\documentclass[conference]{IEEEtran}
\usepackage{cite}
\usepackage{graphicx}
\usepackage{mdwtab}
\usepackage{amsmath}
\usepackage{amssymb}
\usepackage{mathrsfs}
\usepackage{amsfonts}
\usepackage{amsmath}
\usepackage{amsthm}
\usepackage{multicol}
\usepackage{algorithm}
\usepackage{algorithmic}
\usepackage{epsfig}
\usepackage{epstopdf}
\usepackage[utf8]{inputenc}
\usepackage[english]{babel}

\usepackage[pages=some,placement=top]{background}

\renewcommand{\qed}{$\blacksquare$}

\begin{document}
\backgroundsetup{contents=This paper has been submitted to IEEE Transactions on Vehicular Technology.,color=black!100,scale=1.25,opacity=0.7,position={6.75,1.75}}
\BgThispage
\title{Effective Resource Sharing in Mobile-Cell Environments}

\author{\IEEEauthorblockN{Shan Jaffry, Syed Faraz Hasan and Xiang Gui}
\IEEEauthorblockA{School of Engineering and Advanced Technology,\\
Massey University, New Zealand\\
Correspondence: \{S.Jaffry,F.Hasan,X.Gui\}@massey.ac.nz }
}

\maketitle

\begin{abstract}
The mobile users on board vehicles often experience low quality of service due to the vehicular penetration effect, especially at the cell edges. The so-called mobile-cells are installed inside public transport vehicles to serve the commuters. On one end, the mobile-cells have a wireless backhaul connection with the nearest base station, and on the other, they connect wirelessly to the in-vehicle users over access links. 
This paper integrates the mobile-cells within the cellular networks by reusing their sub-channels.
Firstly, this paper proposes an algorithm that allows spectrum sharing for access-link with out-of-vehicle cellular users or MC's backhaul-links. Secondly, it proposes a scheme for controlling the transmit power over the access link to mitigate interference to the backhaul-link, while maintaining high link quality for in-vehicle users.
\end{abstract}

	
\section{Introduction}	

The upcoming Fifth Generation (5G) of cellular technology is expected to boost network performance by supporting a number of new features including multi-layered network architecture \cite{andrews2014will, gupta2015survey}. 
The envisaged heterogeneous network (HetNet) follows a two-layered architecture. The first layer of the this architecture provides coverage to a wider geographical area, and is served by the macrocell eNB (MeNB). The second layer serves the dense cellular regions \cite{cimmino2014role} and is covered by various small-cell eNBs (SeNB), which are fixed low-power base stations. SeNBs can be installed inside stadiums, offices, shopping malls, homes, etc.
More recently, the mobile version of the small-cells have also emerged as wireless hotspots mounted on board vehicles to serve the passengers on trains, subways, or buses etc.
The so-called mobile-cells (MC) provide seamless services to the commuters \cite{jaffry2016making,3GPP_spec,sui2012performance, shanPotenails} and help in minimizing the power loss as the wireless signals penetrate through a vehicle's body. This power loss is termed as the  
vehicular penetration effect (VPE), which can reduce the signal strength by as much as 23 dB \cite{tanghe2008evaluation}. 

\begin{figure}[t]
	\centering
	\includegraphics[width=3.0 in, trim={5.5cm 5.8cm 6.8cm 5.0cm},clip = true]{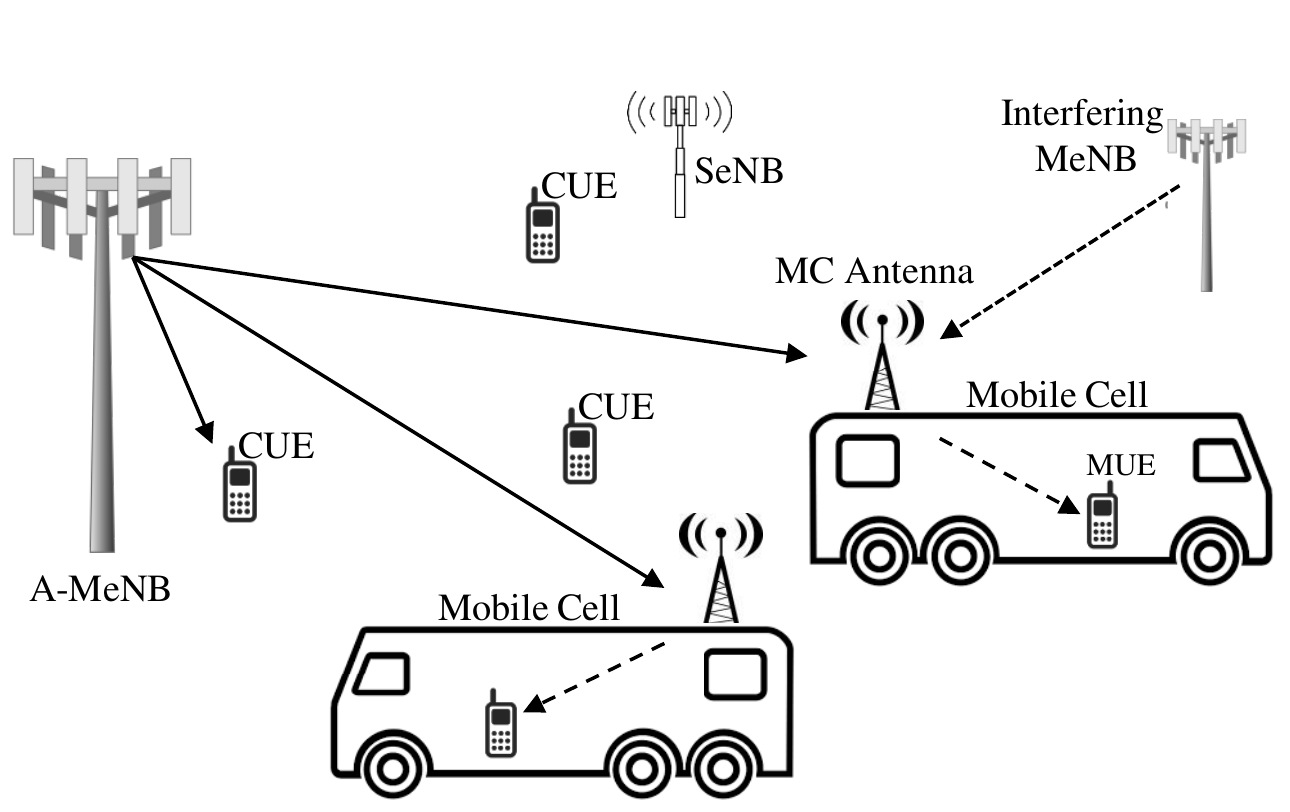}
	\caption{Mobile-cells with active backhaul and access link communication.}
	\label{fig:fig_0}
\end{figure}

An MC comprises of an anchor point that acts as a wireless gateway between the users inside the vehicle and the macrocell base station as shown in Fig. \ref{fig:fig_0}.
An MC typically consists of two antennae: one mounted on top of the vehicles that establishes a connection with the macrocell on the backhaul (BH) link, and the other antenna is housed inside the vehicle to provide coverage to the vehicular users on the access link (AL). 
Therefore, the mobile-cell architecture decouples the MC users (MUE) and the core network. 
The performance analysis of the cellular networks that deploy mobile-cells has been reported in \cite{sui2012performance, jaziri2016offloading, yasuda2015study, sui2013energy}, etc.
It has been shown that the mobile-cells can effectively eliminate VPE, reduce the number of handovers, enhance quality of service (QoS) for commuting users, and increase network throughput. 
In \cite{jiang2015spatio}, the authors have performed experiments to demonstrate that a cache mechanism (such as in mobile-cells) can greatly reduce bandwidth consumption for the public transport users. 
If the employed cache already contains the contents required by a vehicular user, the MC will not need to access the core network.

A significant problem with the mobile-cells is that they may cause severe interference to the macro-cell users 
as demonstrated by the system-level simulations reported in \cite{jaziri2016offloading}. 
It has been argued that the random mobility of the MCs may decrease the overall system throughput due to the interference caused to the fixed cellular layers. 
The authors in \cite{chae2012dynamic} have proposed to use separate bands for in-vehicular and out-of-the-vehicle communication.
This approach maintains the quality of service by eliminating the interference caused due to MC, but at the expense lower spectral efficiency.


\subsection{Related Works}
Spectrally efficient resource allocation algorithms have been presented in 
\cite{jangsher2013resource, jangsher2015resource},
which use graph theory to optimize the power and frequency allocation to the fixed and mobile-cells.
The optimization is meant to minimize interference with the neighboring cells, while maximizing resource utilization. 
Similar technique has been used to assign BH resources for MC in \cite{jangsher2017backhaul}.
Particularly for uplink BH communication, Khan et al. performed analysis and simulation \cite{khan2017outage} to demonstrate that MC can double the coverage of commuting cellular users, along with providing services to the neighboring out-of-vehicle users using cooperative communication. 
Note that the algorithms presented in \cite{chae2012dynamic,jangsher2013resource,jangsher2015resource, jangsher2017backhaul,khan2017outage}
assign resources to any single MC link. However, since MC also has backhaul (BH) and access link (AL) connections, the existing algorithms will lead to spectral inefficiency as they focus on assigning unique resources per wireless link. 

The authors have developed means for spectrum sharing between downlink BH and downlink AL\footnote{In the rest of this paper, we will refer to downlink BH and downlink AL as simply BH and AL, respectively.} for MC in \cite{jaffry2018shared}. 
However, in the regions where MC is close to its associated macrocell eNB (A-MeNB), the BH link may severely interfere with the AL transmission. Similarly, at the cell edge, the communication over the AL may cause interference to BH transmissions. 
Furthermore, BH and AL antennae of an MC are located spatially close to each other. Generally, the transceivers placed in each others' proximity experience large interference if they simultaneously transmit and receive over the same sub-channels \cite{sen2010successive}. This problem can be solved by using techniques such as successive interference cancellation (SIC) \cite{mahmood2016analysing}. 
An SIC enabled receiver is able to decode the stronger signal, subtract it from the original signal stream, and extract the weaker one from the residue \cite{duarte2012experiment}. This technique can further be utilized to enhance the signal-to-interference (SIR) for the stronger signal, thus enabling resource sharing in our scheme.

\subsection{Paper Contribution} Contrary to the existing works, this paper proposes resource sharing scheme for mobile-cell's access-link supporting down link transmissions, which, to the best of authors' knowledge, has not been investigated before. The proposed scheme allows the AL to share resources with either downlink transmission to the out-of-vehicle cellular users or with the backhaul-link of MC. 
The proposed dynamic resource sharing algorithm (DRSA) ensures that AL in the mobile-cells do not require additional sub-channels. 
Secondly, a power control scheme for AL transmissions is also proposed to enhance the quality of received BH signals without compromising on QoS for the in-vehicle users.

The rest of this paper is organized as follows. Section \ref{sec_sys_model} explains the system model. Section \ref{sec_algo} presents the resource sharing algorithm along with mathematical analysis. Section \ref{sec_AL_POWER_control} demonstrates AL power control mechanism. Section \ref{sec_results} discusses the analytical and simulation results. This paper is concluded in Section \ref{sec_conclusion}.

\section{System Model}
\label{sec_sys_model}


\subsection{Network Model}

As we have considered downlink communication for all links, MeNBs and SeNBs are distributed according to homogeneous Poisson point process (PPP) 
$\Phi_M$ and $\Phi_S$ with density $\lambda_M$ and $\lambda_S$ (base-stations/m$^2$) on the Euclidean plane, respectively. Subscript $M$ and $S$ denote the MeNBs and SeNBs in $\Phi_M$ and $\Phi_S$, respectively.
The probability density function of the distance to the neighboring base-stations is given as \cite{andrews2011tractable}:
\begin{equation}
\label{eq_pdf_distance}
f(d) = 2\pi\lambda_{(M,S)} d \exp(-\lambda_{(M,S)}\pi d^2)
\end{equation}
where $\lambda_{(M,S)}$ denotes either of $\lambda_M$ or $\lambda_S$ and $d$ is the distance between the receiving node and its respective base station.
An MC is denoted by $m$ and is located at a random position on the Euclidean plane. 
MCs are considered stationary as in \cite{jangsher2015resource,jangsher2017backhaul,khan2017outage} at a given time instance. 
Each MC is associated with its nearest MeNB (A-MeNB) for BH communication. We do not consider MC communication to SeNBs as MC-to-SeNB BH-links cause large number of handovers due to smaller SeNB coverage \cite{contains2013hetnets, lin2013towards, merwaday2016handover}.

The out-of-vehicle cellular users (CUE) form a set $\mathcal{U}$ and are distributed uniformly over the plane. The individual CUEs are denoted as $u_i$, such that $u_i \in \mathcal{U}$ and  $i\in \{1,2,3...|\mathcal{U}|\}$, where $|.|$ shows the cardinality of the set. For the sake of simplicity and without the loss of generality, we have considered a single MUE per mobile-cell and denote it as $v$.
Each MC has a BH-antenna, which links it to the core network, and a directional AL-antenna ($\tilde{a}$) that is mounted under the roof of the vehicle's body. 
The placement of AL-antenna is such that it enhances the Line of Sight (LoS) component on the AL \cite{rohani2017improving, ni2017self}. 
As for SIC, the performance enhancement is highly dependent on the fact that a common communication control-unit
manages the transmission and reception of signals for BH and AL connections \cite{duarte2012experiment}. 
For simplicity, all concerned notations have been summarized in Table \ref{table_1_param}.

 \begin{table}[t]
	
	\caption{System model parameters} 
	\label{table_1_param}
	\centering
	
	\begin{tabular}{l l }
		\hline
		Symbol & Definition \\
		\hline\hline
		$\mathcal{M},\mathcal{S},\mathcal{U}$ & Set of MeNB, SeNB, CUEs.\\
		$M$,$S$,$\tilde{a}$ & Transmitter for MeNB, SeNB, and AL-antenna.\\
		$\Phi_M, \Phi_S$ & PPP MeNBs, SeNBs.\\
		$\lambda_M, \lambda_S$ & Density of MeNB, Density of SeNB.\\
		$u_i$ & Cellular users registered with MeNB, $u_i \in \mathcal{U}$.\\
		$m$,$u_{\bar{i}}$,$v$ & Receivers for shared MC BH-antenna, CUE, and MUE.\\
		$r_m$& Distance from A-MeNB to MC $m$.\\
		$r_{\tilde{a}v}$ & Maximum distance between AL-antenna and MUE.\\
		$r_{\tilde{a}m}$& Distance between BH-antenna and AL-antenna.\\
		$r_{u_{\bar{i}}}, r_{m u_{\bar{i}}}$ & Distance from A-MeNB to $u_i$, and MC to $u_i$.\\
		$r_{_M},r_{_S}$ & Distance of MC $m$ from interfering MeNBs, and SeNBs.\\
		$r_{_{M'}}$ & Distance of MC $m$ from interfering MeNB, excluding A-MeNB.\\
		$\omega$ & Sub-channel.\\
		$\kappa$ & Small-cell transmission indicator.\\
		$P_M$, $P_S$, $P_{\tilde{a}}$ & Transmit Power of MeNB, SeNB, AL-antenna.\\
		$h^\omega_{t_x,t_r}$&Exponentially distributed fading power for a channel between\\& transmitter-receiver pair $t_x-t_r$ for sub-channel $\omega$.\\
		$\epsilon$ & VPE factor ($0< \epsilon \leq 1$).\\
		$\gamma$ & SIC factor ($0 \leq \gamma \leq 1$).\\

		\hline		
	\end{tabular}
\end{table}

\subsection{Channel Model}
\label{sub_channel_model}

In wireless mobile environments, the channel gain is considered as the product of the large-scale and the small-scale gain \cite{gong2010mobility}.  
All the links to the MC BH-antenna, including MeNB-MC and interfering links, follow quasi-static Rayleigh fading. The large scale attenuation follows standard path loss model i.e. $d^{-\alpha}$, where $d$ is the distance from transmitter to receiver and $\alpha$ is the general path loss exponent. 
Due to the use of directional AL-antenna, AL follows Rician fading where the K-factor determines the strength of the LOS component. The distance of MUE $v$ from AL-antenna is $r_{\tilde{a}v}$. We have used $\alpha_i$ and $\alpha_o$ as the non-LOS and LOS pathloss exponents, respectively. The transmit powers for MeNB, SeNB and MC AL-antenna are represented as $P_M$, $P_S$, and $P_{\tilde{a}}$, respectively.

\subsection{Spectrum Sharing}

\begin{figure}[t]
	\centering
	\includegraphics[width=3.2 in, trim={7.0cm 6cm 5.5cm 7.8cm},clip = true]{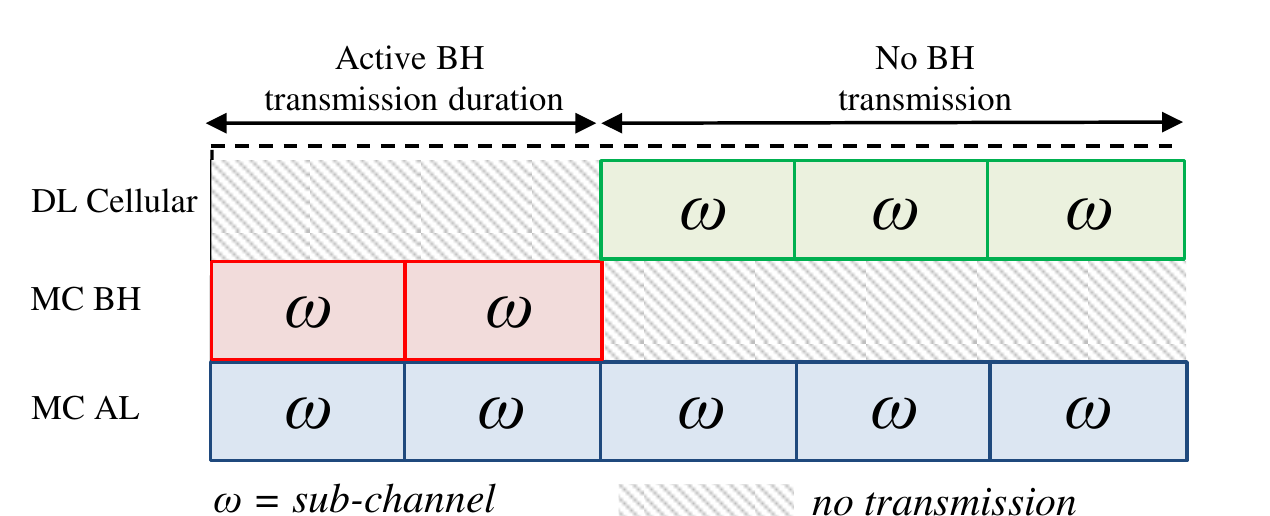}
	\caption{Frequency reuse by different links.}
	\label{fig:fig_0a}
\end{figure}

We consider three links: A-MeNB to MC's BH, AL-antenna to MUE's AL, and A-MeNB to CUE for cellular DL. As shown in Fig. \ref{fig:fig_0a}, the sub-channel $\omega$ is shared by MC AL with either MC BH or cellular link such that $\omega$ is assigned for BH transmission to MC when there is data to be transmitted from A-MeNB to MC. Otherwise, $\omega$ is assigned to out-of-vehicle CUE. The resource sharing algorithm is presented in Section \ref{sec_algo}. 
On the other hand, we define $\kappa$ as the indicator for transmission over sub-channel $\omega$ in the small-cell layer. If all SeNBs utilize sub-channel $\omega$, then $\kappa = 1$. Similarly, when none of the SeNBs transmit over $\omega$, then $\kappa = 0$.
In this study, we have studied the worst-case scenario ($\kappa = 1$, when small-cells are densely populated and are forced to transmit on sub-channel $\omega$), and the best case scenario ($\kappa = 0$, when small-cells are sparsely populated and do not utilize sub-channel $\omega$).

\begin{equation*}
\label{eq_kappa}
\kappa=\begin{cases}
1 & \text{when all SeNBs transmit over RB $\omega$ }.\\
0 & \text{when none of the SeNB transmit over RB $\omega$ }.
\end{cases}
\end{equation*}

\subsection{Signal-to-Interference Ratio}

We have assumed an interference limited environment where noise can be neglected \cite{andrews2011tractable}.  The SIR for BH when $M$ transmit to $m$ over sub-channel $\omega$, is given as:
\begin{equation}
\label{eq_upsilon_1}
	\Upsilon_1(\omega,M\to m) = \frac{P_M r_m^{-\alpha_i} h^\omega_{M,m}}{I_M + I_S + I_{\tilde{a}}\gamma \epsilon  + I'_{\tilde{a}} \epsilon},
\end{equation}
where $r_m$ is the distance from A-MeNB to MC $m$. $I_M = \sum\limits_{M' \in \Phi_M \setminus\{M\}} P_{M'} r_{_{M'}}^{-\alpha_i}h^\omega_{M',m}$ is the total interference from neighboring MeNBs transmitting on the same sub-channel. $I_S = \sum\limits_{S \in \phi_S} P_S r_{_S}^{-\alpha_i}h^\omega_{S,m}$ is the interference from SeNBs and $r_{_{M'}}$, $r_{_S}$ are the distance of MC from interfering MeNB (excluding A-MeNB) and SeNBs, respectively. $I_{\tilde{a}} = P_{\tilde{a}} r_{\tilde{a}m}^{-\alpha_i}h^\omega_{\tilde{a},m}$ is the interference from the AL-antenna of MC $m$, where $r_{\tilde{a}m}$ is the distance between backhaul antenna and AL-antenna. $h^\omega_{t_x,t_r}$ is the exponentially distributed fading power between transmitter ($t_x$) and receiver ($t_r$) for sub-channel $\omega$. Vehicular penetration factor is denoted as $\epsilon$ such that $0 < \epsilon\leq 1$.
The quality of isolation between BH and AL is determined by $\epsilon$. The higher the $\epsilon$, the lower the isolation between the two links. The SIC factor is denoted by $\gamma$ such that $0 \leq \gamma \leq 1$.
Note that the interference from neighboring MC is negligible ($I'_{\tilde{a}} \epsilon \approx 0$) as the AL-antenna acts as the primary source of interference. 
Note that the neighboring mobile-cells residing in the same macrocell are not allocated the same resources. Hence Eq. \ref{eq_upsilon_1} becomes:
\begin{equation}
\Upsilon_1(\omega,M \to m) = \frac{P_M r_m^{-\alpha_i} h^\omega_{M,m}}{I_M + I_S + I_{\tilde{a}} \gamma \epsilon  }.
\end{equation}

The SIR ($\Upsilon_2$) for a cellular user $u_{\bar{i}}$ that uses sub-channel $\omega$ in the downlink is:

\begin{equation}
\label{eq_cellular_user}
\Upsilon_2(\omega, M \to u_{\bar{i}}) = \frac{P_M r_{u_{\bar{i}}}^{-\alpha_i} h_{M,u_{\bar{i}}}^{\omega} }{I_M + I_S + I'_{\tilde{a}}\epsilon}.
\end{equation}
where $r_{u_{\bar{i}}}$ is the distance between $u_{\bar{i}}$ and A-MeNB. $I'_{\tilde{a}}$ is the interference from the AL of the MC $m$ such that $I'_{\tilde{a}} = P_{\tilde{a}} h_{{\tilde{a}},u}^\omega r_{m u_{\bar{i}}}^{-\alpha_i}$, where $r_{m u_{\bar{i}}}$ is the distance between $u_{\bar{i}}$ and $m$.

The SIR for AL at the receiver of user $v$ from AL-antenna $\tilde{a}$ 
on sub-channel $\omega$ is $\Upsilon_3$ which is given as:
\begin{equation}
\label{eq_upsilon3}
\Upsilon_3(\omega,\tilde{a} \to v) = \frac{P_{\tilde{a}} r_{\tilde{a}v}^{-\alpha_o}h^\omega_{\tilde{a},v}}{I_C \epsilon + I'_{\tilde{a}} \epsilon^2},
\end{equation}
where $I_C$ is the interference from cellular transmitters, i.e. $I_C = \sum\limits_{M \in \Phi_M} P_M r_{_M}^{-\alpha_i}h^\omega_{M,v} + \sum\limits_{S \in \phi_S} P_S r_S^{-\alpha_i}h^\omega_{S,v}$ and $r_{_M}$ is the distance of MC $m$ from all interfering MeNBs. We have considered that the AL-antenna is installed at such a position that all users are within the LOS range of the transmitter. Note that the interference from neighboring MCs' AL-antennas is negligible. Hence for all interfering MC AL-antenna $\tilde{a}'$, $ I'_{\tilde{a}}\epsilon^2 \approx 0.$ Hence Eq. \ref{eq_upsilon3} becomes:
\begin{equation}
\Upsilon_3(\omega,\tilde{a} \to v) = \frac{P_{\tilde{a}} r_{\tilde{a}v}^{-\alpha_o}h^\omega_{{\tilde{a}},v}}{I_C \epsilon}.
\end{equation}

\section{Dynamic Resource Sharing Algorithm}
\label{sec_algo}

In this section we present Dynamic Resource Sharing Algorithm (DRSA). Note that when there is no data to be transmitted from A-MeNB to MC, $\omega$ is assigned to CUE $u_{\bar{i}}$, which is selected from the CUE set $\mathcal{U}$ based on the distance between MC $m$ and the user $u_{\bar{i}}$, as shown in Algorithm I.

\begin{algorithm}[t]
	\label{algo_11}
	\begin{algorithmic}
		
		\caption[short]{DRSA Algorithm}
		\STATE { A-MeNB : Associated MeNB.\\
			$\omega$ : Sub-channel.\\
			$r_{u_i}$ : Distance between A-MeNB and CUEs ($u_i$).\\
			$\mathcal{U}$ : Number of cellular users registers with A-MeNB.\\
			$r_{m u_i}$ :  Distance between CUE ($u_i$) and MC ($m$).\\
			$P_{\tilde{a}}$ : Transmit power for MC AL-Antenna.\\
			$\bar{t}$ : Unit of time.
		}
		\STATE
		\FORALL {$\bar{t}>0$}
		\STATE A-MeNB knows MC-location at time $\bar{t}$.
		\STATE A-MeNB knows  $r_{u_i}$ and $r_{m u_i}$.
		\IF{\ there exist data for BH transmission} 
		\STATE  {A-MeNB assigns $\omega$ to MC BH.} 
		\STATE  {MC AL shares $\omega$ with MC BH.}
		\ELSE 
		\STATE {A-MeNB assigns $\omega$ to CUE ($u_{\bar{i}}$) such that:} 
		\FORALL {$i \leq |\mathcal{U}|$}
		\STATE {\ \ \ $\min \text{sort}$ \Large$\frac{r_{u_i}}{r_{m u_i}}$\normalsize $\forall \ r_{u_i} | r_{m u_i}$}
		\ENDFOR
		\STATE {MC AL shares $\omega$ with CUE ($u_{\bar{i}}$) DL such that:} 
		\STATE {\ \ \  i. \Large$\frac{r_{u_{\bar{i}}}}{r_{m u_{\bar{i}}}}$\normalsize $< 1 $}
		\STATE {\ \ \  ii. \Large$\frac{r_{u_{\bar{i}}}}{r_{m u_{\bar{i}}}} < \frac{r_{u_i}}{r_{m u_i}}$ \normalsize$\forall \ \bar{i} \not= i$, $i\in \{1,2,3,...|\mathcal{U}|\}$}
		\ENDIF
		\ENDFOR
	\end{algorithmic}
\end{algorithm}

Algorithm I outlines the scheme that allows sharing of sub-channel $\omega$ over multiple links. In the following, we will analyze the performance of the proposed resource sharing algorithm with specific regard to the quality of service. 
The parameter of interest is success probability ($p$), which depends on the SIR at the receiver. The received SIR should be greater than the threshold $\theta$ for a transmission to be successful.
Mathematically, success probability for a link between transmitter $t_x$ and receiver $t_r$ over sub-channel $\omega$ can be represented as $p = \mathbb{P}[\Upsilon(\omega,t_x \to t_r) > \theta]$, where $\mathbb{P}[.]$ represents the probability of an event. Success probabilities of the individual links are covered in the following:

\subsection*{Theorem 1: Success Probability for Backhaul link}
\label{theorem_1}
Success probability for backhaul links is calculated for the following two cases:

\subsubsection{When $\kappa = 1$}

\begin{equation}
\label{eq_p1_0}
p_{BH} = \int\limits_{0}^{1} \frac{1}{\varpi^2}\exp\Bigg\{-\Bigg(\frac{1}{\varpi}-1\Bigg)\mathcal{Z} \Bigg\} \times \frac{1}{1 + \mathcal{Y}_1(\frac{1}{\varpi}-1)^2}d\varpi.
\end{equation}
where $\mathcal{Z} = \sqrt{\theta}\big(\pi/2 -\arctan(1/\sqrt{\theta})\big) + \frac{\lambda_S}{\lambda_M}\frac{\pi}{2}\sqrt{\frac{P_S}{P_M}\theta} + 1$ and 

\begin{equation}
\label{eq_success_link1}
\mathcal{Y}_1 = \frac{\gamma \epsilon \theta P_{\tilde{a}}}{P_M\pi^2\lambda_M^2r_{\tilde{a}m}^4}.
\end{equation}

\subsubsection{When $\kappa = 0$}
\label{sub_sec_kappa0}
\begin{equation}
\label{eq_p1_1}
p_{BH} = 2\pi \lambda_M \int_{0}^{1}  \frac{1}{z^2}\Big(\frac{1}{z}-1\Big)\frac{\exp{\Big\{-\mathcal{Z}' \Big(\frac{1}{z} - 1\Big)^2\Big\}}}{1 + \mathcal{Y}_2(\frac{1}{z}-1)^{4}}  dz.
\end{equation}

where $\mathcal{Z}' = \pi \lambda_M (1 + \sqrt{\theta}(\pi/2 - \tan^{-1}(1/\sqrt{\theta})))$ and

\begin{equation}
\label{eq_success_link2}
	\mathcal{Y}_2 = \gamma \epsilon \theta  \frac{P_{\tilde{a}}}{P_M r_{\tilde{a}m}^{4}}.
\end{equation}

Proof: See Appendix \ref{app_success_DL_BH}.\\

The variables $\mathcal{Y}_1$ and $\mathcal{Y}_2$ are termed as BH success-link parameters, which help in increasing the success probability for the BH link.
They have been expressed in terms of $P_{\tilde{a}}$, which is derived in the next section. 
The success-link parameters determine the relationship between BH success probability and transmit power of AL-antenna.
Also note that Eq. \ref{eq_p1_0} and Eq. \ref{eq_p1_1} are not the closed form expressions and are difficult to solve analytically. Hence, they are evaluated numerically and results are presented in Section \ref{sec_results}.

\subsection*{Theorem 2: Success Probability of Shared Cellular DL}
	\label{theorem_2}
	
	In case there is no BH communication for MC, MUE $v$ shares the sub-channel $\omega$ with a cellular user $u_{\bar{i}}$. The success probability for the shared DL cellular transmission is given as:	
	\begin{equation}
	\label{eq_p3_final}
	p_{DL} = \frac{\exp\Big\{ \frac{-\pi r_{u_{\bar{i}}}^2 }{2}\sqrt{\frac{\theta}{P_M}}\Big(\lambda_M\sqrt{{P_M}} + \kappa\lambda_S\sqrt{{P_S}}\Big)\Big\}}{1 + \frac{\theta\epsilon}{P_M}\Big(\frac{r_{u_{\bar{i}}}}{r_{m u_{\bar{i}}}}\Big)^4}.
	\end{equation}

Proof: See Appendix \ref{app_success_DL_CL}.\\

Note that in Eq. \ref{eq_p3_final}, $p_{DL}$ depends on the distance of CUE $u_{\bar{i}}$ from A-MeNB ($r_{u_{\bar{i}}}$), and on the distance between $u_{\bar{i}}$ from MC $m$ ($r_{m u_{\bar{i}}}$).

\subsection*{Theorem 3: Success Probability for Access-Link}
\label{theorem_3}
The success probability for access-link is given as: 

\begin{multline}
\label{eq_p2_final2}
p_{AL} = \sum\limits_{j=0}^{J} \sum\limits_{m=0}^{j} \frac{K^j (-\theta)^{j-m}}{e^K j!(j-m)!}\sum\limits_{q=0}^{Q} \frac{(-1)^q \Omega_{\kappa}^q}{q!}.\frac{1}{P_{\tilde{a}}^q}\\\theta^{\frac{2q}{\alpha_i} - (j-m)}\Psi_{(2q/\alpha_i,j-m)}.
\end{multline}
where  $\Psi_{(2q/\alpha_i,j-m)} = \frac{\Gamma(2q/\alpha_i+1)}{\Gamma(2q/\alpha_i-(j-m) + 1)}$
and $\Gamma(.)$ is the Gamma function. $K$ is the Rician K-factor and $J$, $Q$ are the upper limits for index parameter $j$, $q$ respectively. The numerical values of $J,Q$ are given in Section \ref{sec_results}, and
\begin{multline*} 
\Omega_{\kappa} = \pi (\gamma \epsilon r_{\tilde{a}v}^{\alpha_o})^{2/\alpha_i} \Bigg(\lambda_M P_M^{2/\alpha_i} + \kappa\lambda_S P_S^{2/\alpha_i}\Bigg) \beta(\alpha_i).
\end{multline*}
where $\beta(\alpha) = \frac{2\pi/\alpha}{\sin(2\pi/\alpha)}$.

Proof : See Appendix \ref{app_success_AL} 

\section{Transmit Power control over Access Link}
\label{sec_AL_POWER_control}

The signals received on the mobile-cell's BH link experience high interference from the AL-antenna transmission that is in close vicinity to the BH-antenna. On the other hand, the transmission quality over AL is affected by the neighboring cellular transmitters (MeNBs and SeNBs). However, since 
it uses directional antennae, AL can yield high success probability even with smaller transmit power. 
We therefore propose that the MC adaptively controls the transmit power of the AL-antenna ($P_{\tilde{a}}$) to enhance the reception quality over BH link. The increased reception quality should not compromise the quality of service for the in-vehicle cellular users. The proposed adaptive power control scheme can be moderated by the MC's communication control unit \cite{duarte2012experiment}.

Considering $\alpha_i = 4$ for AL success probability in Eq. \ref{eq_p2_final_app}, and using the property $e^{(.)} = \sum\limits_{q=0}^{\infty}\frac{(.)^q}{q!}$, $p_{AL}$ can be written as:
\begin{multline}
p_{AL} = \exp\Big(\frac{-\Omega_{\kappa}\sqrt{\theta}}{P_{\tilde{a}}}\Big)\sum\limits_{j=0}^{\infty} \sum\limits_{m=0}^{j} \frac{K^j (-\theta)^{j-m}}{e^K j!(j-m)!}\sum\limits_{q=0}^{\infty} \theta^{- (j-m)}\\\Psi_{(2q/\alpha_i,j-m)},
\end{multline}
where $\Psi_{(2q/\alpha_i,j-m)}$ and $\Omega_{\kappa}$ are given in Theorem 3.
Then, AL antenna's transmit power will be:
\begin{equation}
\label{eq_po}
P_{\tilde{a}} = \frac{\Omega_{\kappa}\sqrt{\theta}}{\log(\Xi)- \log(p_{AL})}.
\end{equation}
where
\begin{equation*}
\Xi = \sum^{\infty}_{j=0}\sum^{j}_{m=0}\frac{K^j (-1)^{j-m}}{e^K j! (j-m)!}\sum^{\infty}_{q=0}\Psi_{(2q/\alpha_i,j-m)}.
\end{equation*} 

Note that $P_{\tilde{a}}$ in Eq. \ref{eq_po} is the function of AL's success probability ($p_{AL}$), which can be determined by MC communication control-unit based on transmission requirement for the in-vehicle users. Other parameters such as K-factor for LoS transmission, $\gamma, \epsilon$ are assumed to be pre-determined by the MC communication control-unit.
The parameters such as $\lambda_M$ and $\lambda_S$ can be broadcast by 
A-MeNB \cite{molina2017lte, jaffry2018neighbourhood}. We have assumed fixed transmission power for macrocell and small-cell layers. 

For BH, the expressions of the success probability for the case when $\kappa = 1$ and when $\kappa = 0$ are given as Eq. \ref{eq_p1_0} and Eq. \ref{eq_p1_1}, respectively. Using AL power control, the BH success-link parameters ($\mathcal{Y}_1$ and $\mathcal{Y}_2$ as given in Eq. \ref{eq_success_link1} and Eq. \ref{eq_success_link2}) can be modified by inserting $P_{\tilde{a}}$ from Eq. \ref{eq_po} as:
	  
	\subsubsection*{When $\kappa = 1$}

	\begin{equation}
	\label{eq_Y1}
	\mathcal{Y}_1 = \Bigg(\frac{\gamma\theta \epsilon }{r_{\tilde{a}m}^4P_M\pi^2\lambda_M^2}\Bigg)\Bigg( \frac{\Omega_{\kappa}\sqrt{\theta}}{\log(\Xi)-\log(p_{AL})}\Bigg).
	\end{equation}
	
	\subsubsection*{When $\kappa = 0$}
	
	\begin{equation}
	\label{eq_Y2}
	\mathcal{Y}_2 =   \Bigg(\frac{\gamma \epsilon \theta}{P_M r_{\tilde{a}m}^{4}}\Bigg)\Bigg( \frac{\Omega_{\kappa}\sqrt{\theta}}{\log(\Xi)-\log(p_{AL})}\Bigg).
	\end{equation}
\section{Ergodic Rate and Sub-Channel Reuse}

In this section, we first measure the ergodic rate per sub-channel in nats/sec/Hz (1 nat $\approx$ 1.443 bits) for each link. The analysis of sub-channel reuse following the proposed algorithm is also presented in this Section.

The ergodic rate for BH ($T_{BH}$) is derived as Eq. \ref{eq_Rates_3} \cite{andrews2011tractable}, where \large$\mathcal{F} = \frac{P_{\tilde{a}} \gamma \epsilon}{P_M r_{\tilde{a}m}^4}$\normalsize. 

\begin{figure*}[t]
	\hrule
	\begin{equation}
	\label{eq_Rates_3}
	T_{BH} = \int\limits_{g=0}^{g=1} \int\limits_{\sigma=0}^{\sigma=1} \frac{\exp\bigg\{-\bigg(\frac{1}{\sigma}-1\bigg)\bigg[ 1 + \sqrt{e^{\frac{1}{g} - 1}-1}\bigg(\frac{\pi}{2} - \tan^{-1}\frac{1}{\sqrt{\exp(1/g-1) - 1}} + \frac{\kappa \lambda_S}{2\lambda_M}\sqrt{\frac{P_S}{P_M}}\bigg)\bigg] \bigg\} }{g^2 \sigma^2 \bigg[1 + \big(\frac{1}{\sigma}-1\big)^2\bigg( \frac{\mathcal{F}\exp(1/g -1) - 1}{\lambda_M^2 \pi^2} \bigg)\bigg]}\ d\sigma\ dg.
	\end{equation}
	\hrule
\end{figure*} \normalsize

On the other hand, the ergodic rate of the cellular downlink is given as:

\begin{equation}
\label{eq_erg_cellular_DL}
T_{DL} = \int_0^1 \frac{\exp\Big(-\tilde{A}\sqrt{e^{\frac{1}{\tilde{g}}-1}-1} \Big)}{\tilde{g}^2 (1 + \tilde{B} (e^{\frac{1}{\tilde{g}}-1}-1))} d\tilde{g}.
\end{equation} \normalsize
where $\tilde{A} = \frac{\pi r_{u_{\bar{i}}}^2(\lambda_M + \lambda_S\kappa\sqrt{\frac{P_S}{P_M}})}{2}$ and $\tilde{B} = \frac{1}{P_M}\Big(\frac{r_{u_{\bar{i}}}}{r_{m u_{\bar{i}}}}\Big)^2 $\\



And finally, AL's ergodic rate is given as:
\begin{multline}
\label{eq_Ergodicrate_AL}
\small T_{AL} = \int\limits_{0}^{1}\frac{1}{g^2} \sum_{j=0}^{\infty}\sum_{m=0}^{j}\sum_{q=1}^{\infty}\\\frac{K^j(-1)^{j-m+q}(e^{\frac{1}{g}-1}-1)^{2q/\alpha}}{e^Kj!(j-m)!}\frac{\Omega_k^q}{q!} \Psi_{(2q/\alpha_i,j-m)}\ dg.
\end{multline} \normalsize 
where $\Psi_{(2q/\alpha_i,j-m)}$ is defined in Theorem 3.\\

Complete derivations of $T_{BH}$, $T_{DL}$ and $T_{AL}$ are given in Appendix D.
We use numerical integration to solve Eq. \ref{eq_Rates_3}, \ref{eq_erg_cellular_DL} and  \ref{eq_Ergodicrate_AL}, and report the results in Section \ref{sec_results}.

\subsection*{Sub-channel Reuse}

The sub-channel reuse factor ($\mathcal{Q}_\omega$) is defined as the number of times a sub-channel $\omega$ is used under the coverage of an MeNB. Note that since AL shares the resource with either BH or DL cellular communication, hence $\mathcal{Q}_\omega \geq 2$. 
However, if the small-cells also transmit on the same sub-channels, $\mathcal{Q}_\omega$ becomes:
\begin{equation}
\label{eq_final_RUF}
\mathcal{Q}_\omega = \max (2 , 2 +\sum\limits_{\eta=0}^{\lambda_S |B|} \kappa_\eta).
\end{equation}
where $\kappa_\eta$ is an indicator function such that $\kappa_\eta \in \{0,1\}$ is $1$ when small-cell $\eta$ transmits on the same frequency as MC's AL, and $0$ when the small-cell is silent. $|B|$ is the area-bound (in square meter) around an A-MeNB.

We can see from Eq. \ref{eq_final_RUF} that the sub-channel reuse factor has a linear relationship with the density of the small-cells ($\lambda_S$) as well as with the indicator function $\kappa_\eta$. Intuitively, it can be seen that the reuse-factor will be 2 when no small-cell transmit on the sub-channels shared by MC.
\begin{table}[t]
	
	\caption{Simulation Parameters} 
	\label{table_sim_param}
	\centering
	
	\begin{tabular}{l l }
		\hline
		Symbol & Definition \\
		\hline  			
		Simulation Runs & 10,000\\
		Simulation Area & $40 \times 40$ sq. km\\
		Transmit powers ($P_M, P_{\tilde{a}}$) & 45, 0 dBm\\
		Max. AL-antenna $\leftrightarrow$ MUE distance& 8 meters\\
		BH $\leftrightarrow$ AL antennae distance & 5 meters\\
		$\alpha_i$ / $\alpha_o$  & 4/3.5 \\
		$r_{u_{\bar{i}}}$ & 50 meters\\
		$J,Q$ & 70\\
		\hline		
	\end{tabular}
\end{table}

\begin{figure}[t]
	\centering
		\includegraphics[width=3.25 in, trim={5.0cm 3.0cm 4.5cm 4.0cm},clip = true]{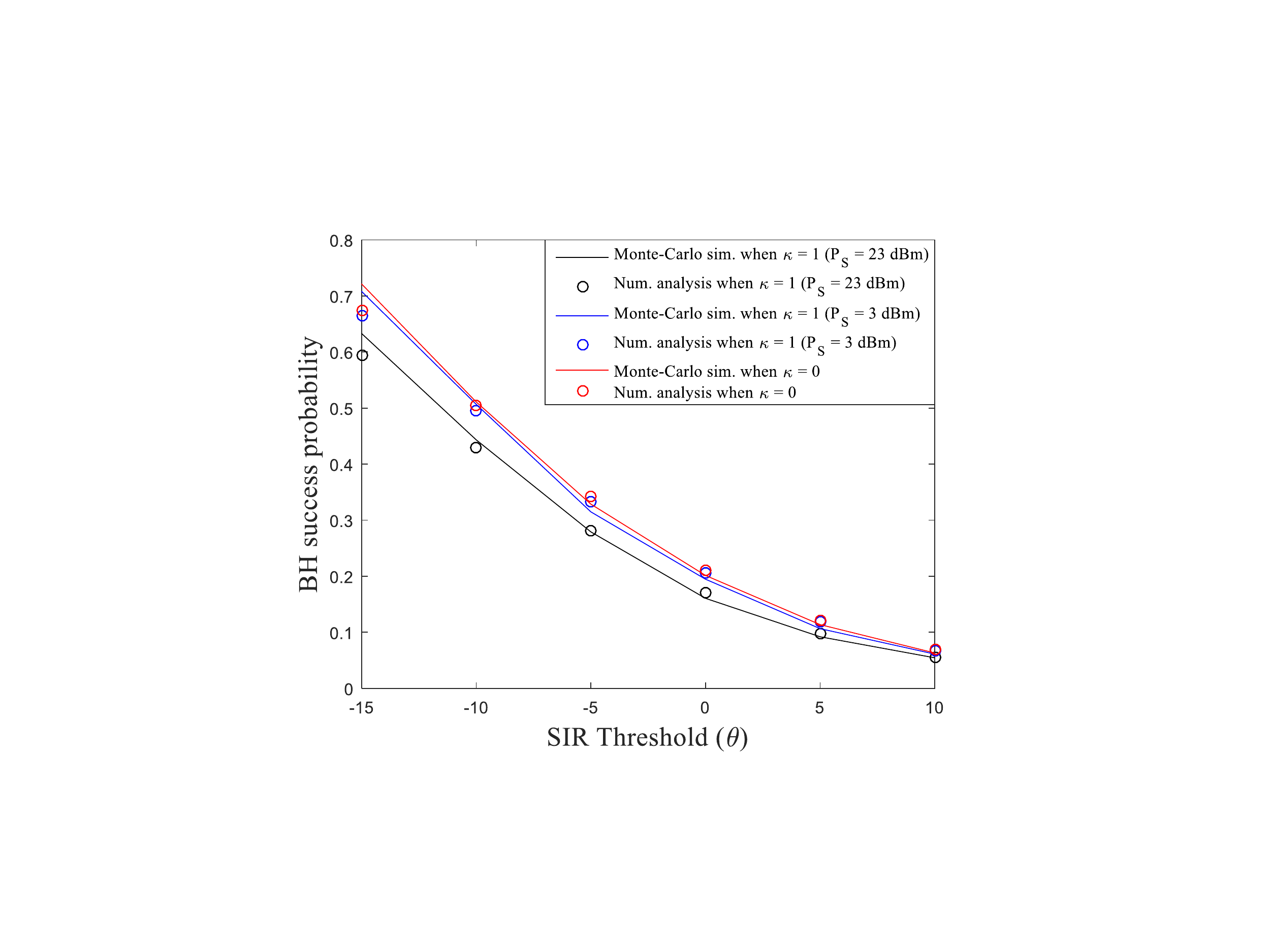}
	\caption{BH link Success Probability ($\lambda_M = 2\times10^{-6}, \lambda_S = 10\times\lambda_M$, $\epsilon = 0.1$, $\gamma = 1$ i.e. No SIC)}
	\label{fig:fig_3b}
\end{figure}

\begin{figure}[t]
	\centering
	\includegraphics[width=3.25 in, trim={4.5cm 4.2cm 4.7cm 4.5cm},clip = true]{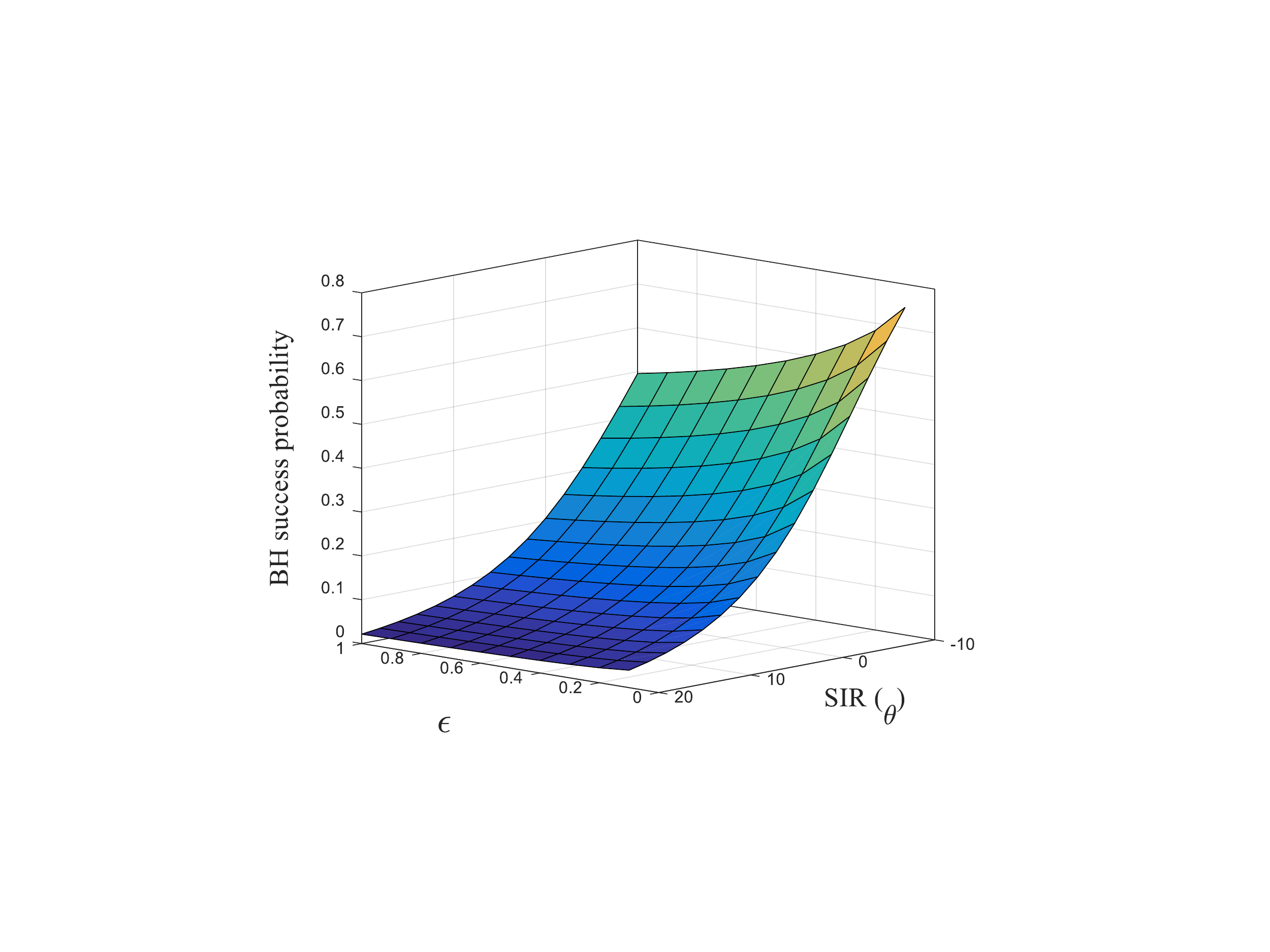}
	\caption{Dependence of BH Success probability on SIR and $\epsilon$ ($\lambda_M = 2\times10^{-6}, \kappa = 1, P_S = 3 dB$)}
	\label{fig:fig_3c}
\end{figure}

\begin{figure}[t]
	\centering
	\includegraphics[width=3.25 in, trim={4.2cm 3.0cm 4.0cm 4.0cm},clip = true]{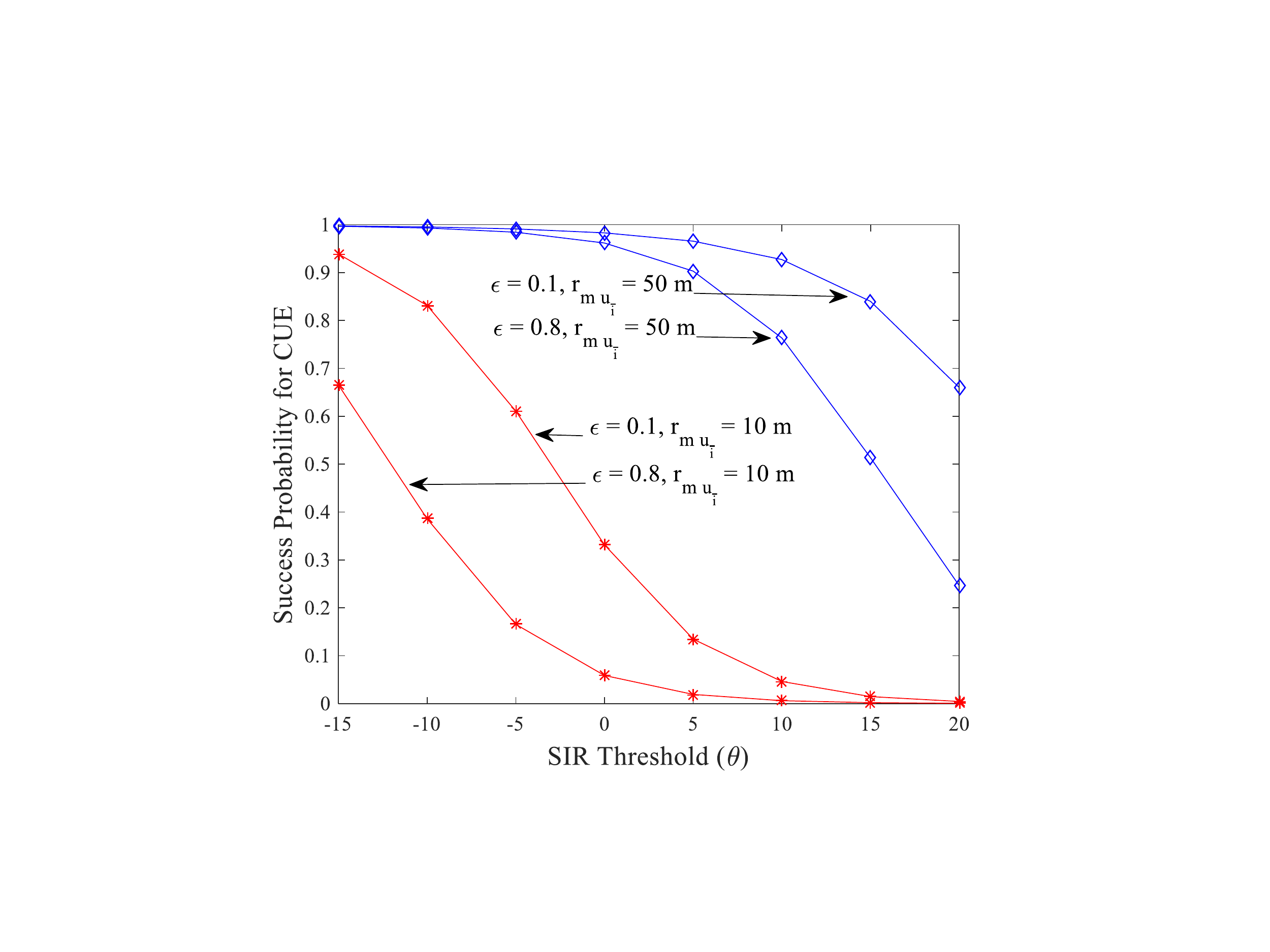}
	\caption{Dependence of DL Success probability on SIR and $\epsilon$ ($\lambda_M = 4\times10^{-6}, \lambda_S = 10\times\lambda_M, r_{u_{\bar{i}}} = 50m, \kappa = 1$)}
	\label{fig:fig_3d}
\end{figure}

\section{Simulation and results}
\label{sec_results}

In this section, we present the evaluation for the proposed DRSA algorithm using Monte-Carlo simulations. 
The total area considered for simulations is 40 Km $\times$ 40 Km. We averaged the results for 10,000 realizations for each simulation. 
The simulation results reported below have been validated using mathematical analysis.
Table \ref{table_sim_param} contains the concerned simulation parameters. 

\subsection{Backhaul Link Performance Evaluation}

Fig. \ref{fig:fig_3b} shows the success probability for the BH link for three cases: $\kappa = 0$, $\kappa = 1$ and $P_S = 23$ dBm, and when $\kappa = 1$ and $P_S = 3$ dBm. It can be seen that the difference between the BH success probability when small-cells transmit with low power ($P_S = 3$ dBm), and when $\kappa = 0$ is marginal. On the other hand, when the small-cells transmit with relatively higher power, the success probability is lower compared to the other two cases. This is intuitive because the main source of interference is the AL-antenna, which is located closer to the BH-antenna. The slight difference between the analytical and simulation result (see Fig. \ref{fig:fig_3b}) is because we have considered numerical approximations in Eq. \ref{eq_p1_0} and Eq. \ref{eq_p1_1}, as mentioned in Section \ref{sec_algo}. Fig. \ref{fig:fig_3c} demonstrates the combined effect of SIR Threshold $(\theta)$ and the penetration factor ($\epsilon$) on the success probability of the BH link. It is evident from the figure that success probability for BH will increase with higher VPE (i.e. lower values of $\epsilon$), especially for lower SIR thresholds.

\begin{figure}[t]
	\centering
	\includegraphics[width=3.25 in, trim={5.0cm 3.2cm 4.0cm 3.7cm},clip = true]{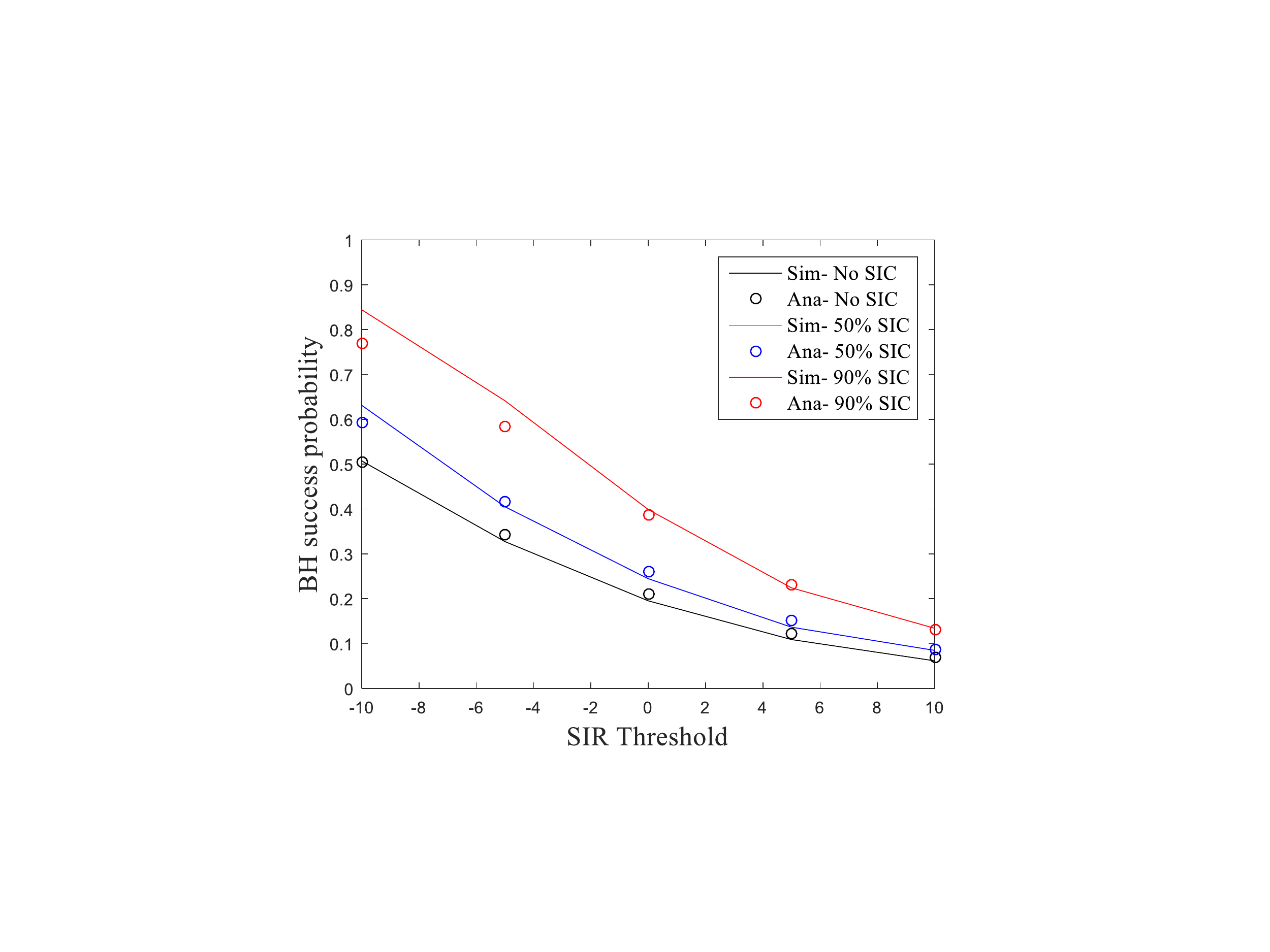}
	\caption{BH Success probability for different Penetration-factors ($\epsilon = 0.1$, $\lambda_M = 2\times10^{-6}, \kappa = 0$) }
	\label{fig:fig_3e0}
\end{figure}

\begin{figure}[t]
	\centering
	\includegraphics[width=3.25 in, trim={4.6cm 3.0cm 4.0cm 3.8cm},clip = true]{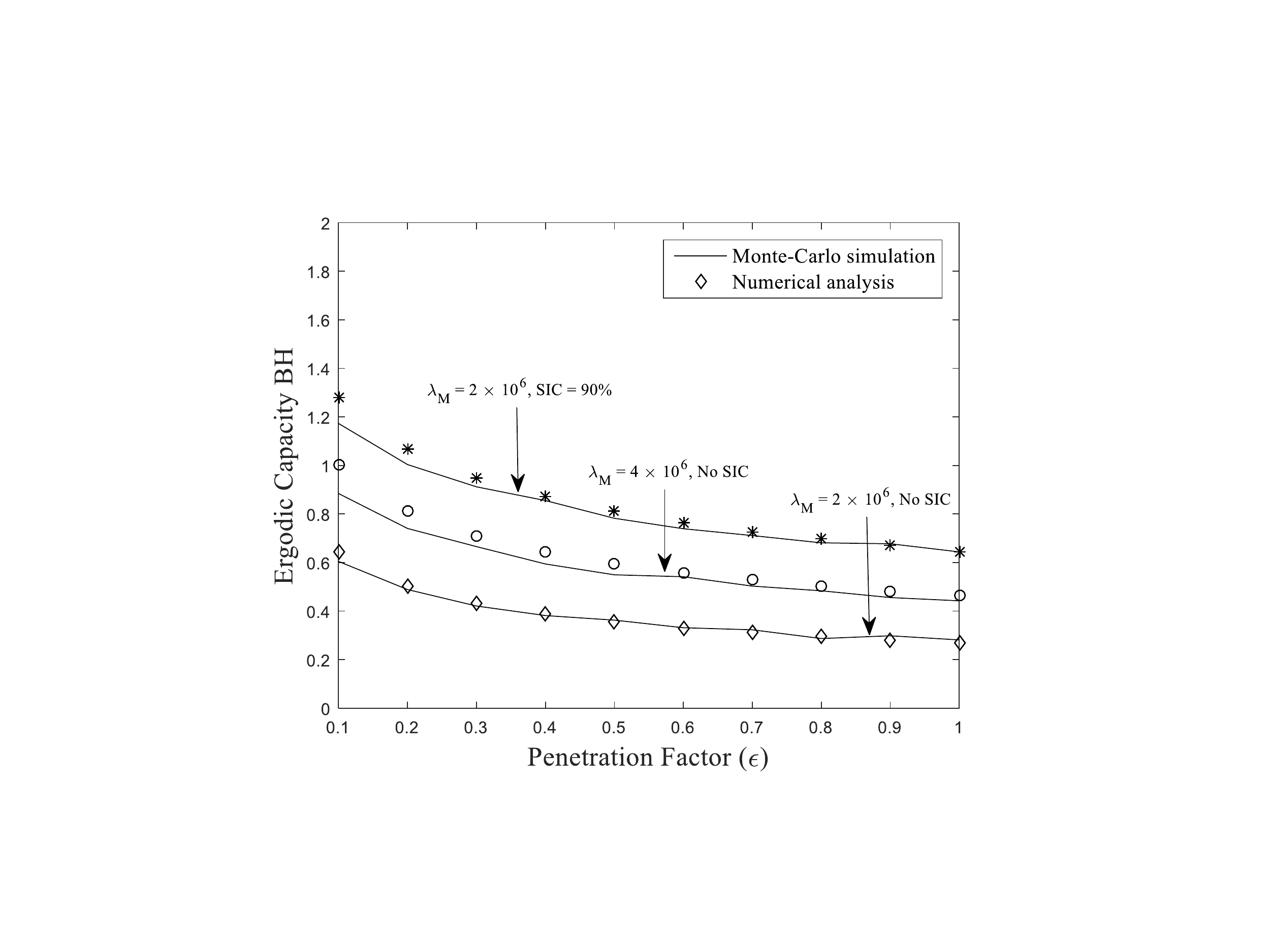}
	\caption{Ergodic rate for BH link in nats/sec/Hz }
	\label{fig:fig_3e1}
\end{figure}

Fig. \ref{fig:fig_3d} presents the success probability for the downlink cellular link for out-of-vehicle user $u_{\bar{i}}$. It can be seen that the link quality improves significantly as the distance between the mobile-cell and out-of-vehicle user increases, enabling efficient resource sharing between AL and the cellular downlink. 
The impact of SIC on the success probability of BH has been shown in Fig. \ref{fig:fig_3e0}. 
It is obvious that the success probability can be higher for an ideal SIC-enabled system. 
On the contrary, with no SIC, BH link quality degrades drastically, especially for higher SIR threshold values.
\linebreak Fig. \ref{fig:fig_3e1} shows the ergodic capacity of the BH link. It is evident from the figure that the ergodic capacity per sec/Hz will increase if SIC is employed. Even for the sparse macrocell deployment (e.g. when $\lambda_M = 2\times10^{-6}$), the capacity of BH link with $90\%$ SIC is higher than the BH capacity when the macrocells are densely deployed (e.g. $\lambda_M = 4\times10^{-6}$) and there is no SIC at BH.
 
\subsection{Access Link Performance Evaluation}

\begin{figure}[t]
	\centering
	\includegraphics[width=3 in, trim={5.0cm 3.3cm 5.8cm 4.0cm},clip = true]{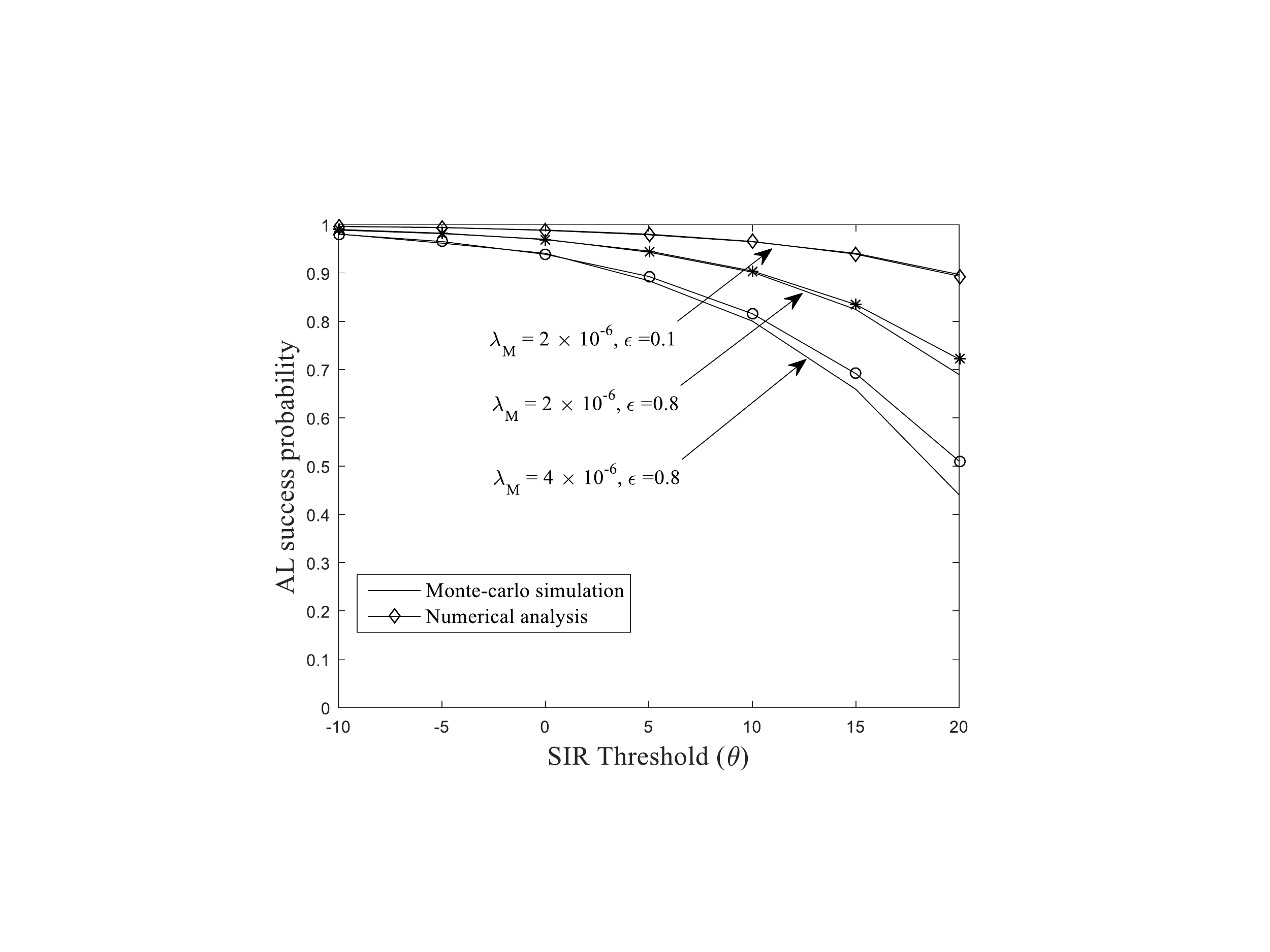}
	\caption{AL Success Probability ($\kappa = 0$, $\gamma = 1$ i.e. No SIC)}
	\label{fig:fig_4}
\end{figure}

Fig. \ref{fig:fig_4} demonstrates the performance analysis of AL. Note that due to the use of directional antennae, the quality of AL is high even for the higher values of $\theta$ and $\epsilon$. 
AL success probability approaches 1 when high VPE prevents interference from BH (e.g. $\epsilon = 0.1$). 
The Rician K-factor is assumed to be 2 dB, which means that there exists a nominally dominant LOS component as compared to the multipath components.
Also note from Fig. \ref{fig:fig_4} that at very high SIR thresholds, success probability over AL deteriorates for poorly isolated MC structure (e.g. $\epsilon = 0.8$). 
On the contrary, if we assume a strong LOS, then the success probability for AL would be very high, even for poorly-isolated scenarios. However, in practical scenarios, $K$ does not attain high values (e.g. $K\geq10$, etc).

\begin{figure}[t]
	\centering
	\includegraphics[width=3.25 in, trim={5.0cm 3.6cm 5.5cm 4.5cm},clip = true]{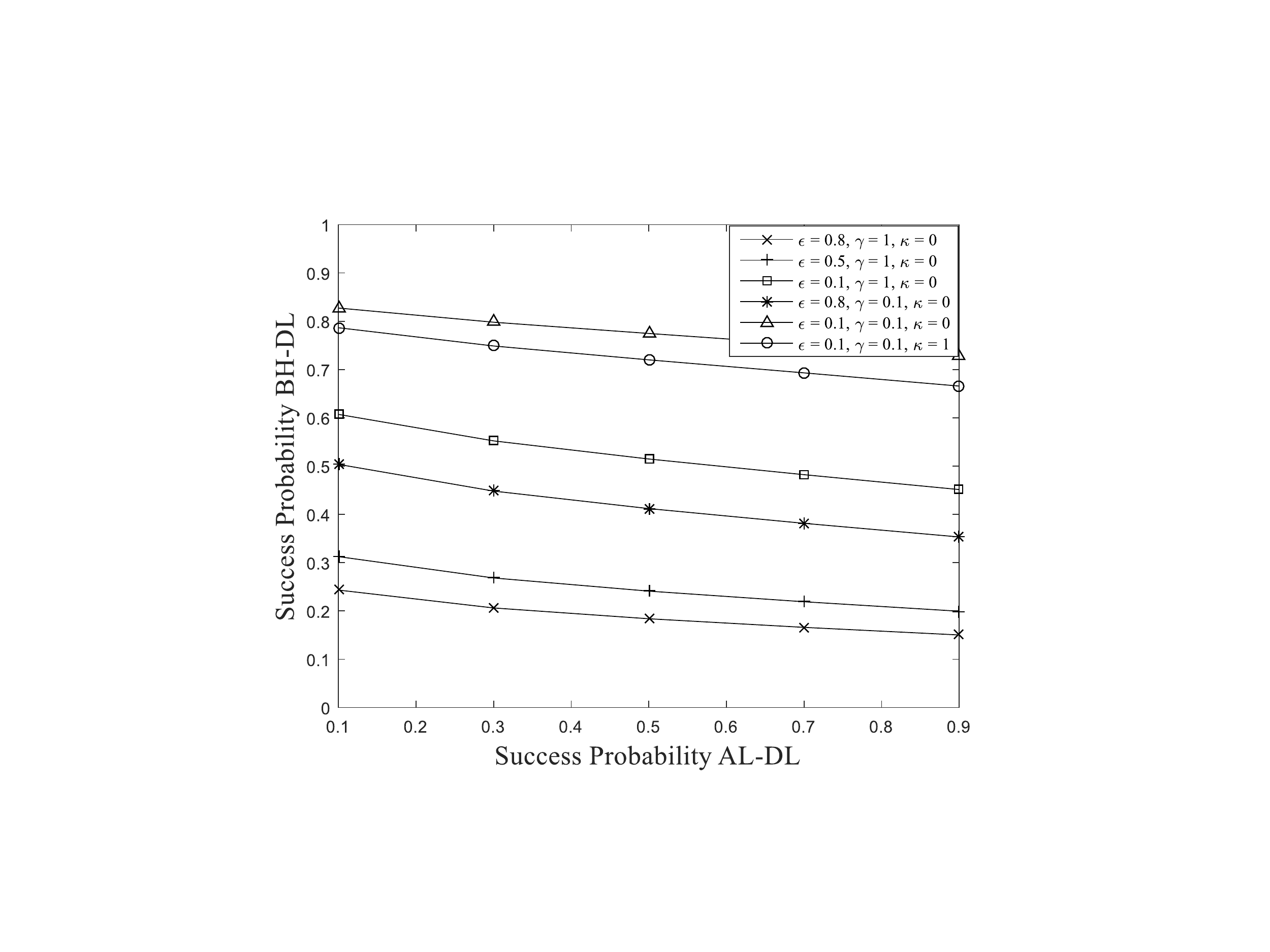}
	\caption{BH Success Probability vs AL Success Probability ($\lambda_M = 2\times 10^{-6}, \kappa = 0, \theta = -10$  dB)}
	\label{fig:fig_5}
\end{figure}

Finally, Fig. \ref{fig:fig_5} shows the relationship between the success probability of AL and BH. It can be observed that with high MC-structural isolation, alongside accurate SIC suppression, AL and BH links can offer highly successful transmissions.
It is interesting to note that the success probability for AL depends, among other factors, on the transmissions power of the AL antenna (refer to Eq. \ref{eq_po}). It means that based on the nature of the transmission required by AL, transmit power of the AL-antenna ($P_{\tilde{a}}$) may be adjusted dynamically, which will consequently affect the BH link.

\section{Conclusion}
\label{sec_conclusion}
In this research, we have catered a serious challenge posed by integrating MC with cellular network. Following conventional spectrum allocation schemes, distinct resources would be required for wireless backhaul and access links for MC. Such schemes would reduce network spectral efficiency. We proposed dynamic resource sharing algorithm to ensure that AL shares sub-channel either with MC's BH link or out-of-vehicle cellular user. VPE was exploited along with using SIC technique to reduce interference between the shared links. It was observed that for lower values of penetration factor ($\epsilon$), higher success probabilities for BH, AL, and cellular link are achieved. Further results demonstrated that nearly 100\% increase in the ergodic rate of BH link is achieved by employing SIC. The access-link power-control was also proposed to reduce the interference to the backhaul link. It was shown that with high penetration factor and efficient SIC, BH and AL attain high success probability, simultaneously. 

Resource management for sidehaul links between mobile-cells is the focus of our on-going research. 

\appendices
\section{Downlink Backhaul Success Probability}
\label{app_success_DL_BH}

Let $p_{BH}$ be the Success Probability for BH

\begin{equation}
\label{main_eq_p1}
p_{BH} = \mathbb{E}_{r_m}\  \big[\mathbb{P}[\Upsilon_1(\omega,M\to m) > \theta \ |\ r_m]\big],
\end{equation}
where expectation $\mathbb{E}[.]$ is with respect to the location of MC around A-MeNB. 
\begin{equation}
\label{eq_p1_beginning}
p_{BH} = \int_{r_m>0} \mathbb{P}[\Upsilon_1(\omega,M\to m) > \theta \ |\ r_m] \ f_c(r_m) dr_m,
\end{equation}

Since $h^\omega_{M,m} \sim \exp(1)$, we use the complimentary cumulative distribution (CCDF) of $h^\omega_{M,m}$ to re-write Eq. \ref{eq_p1_beginning} as:

\begin{multline}
= \mathbb{E} \Big[\int_{r_m>0} \exp\Big(- \frac{\theta r_m^{\alpha_i}}{P_M} (I_M + I_S + \gamma I_{\tilde{a}}\epsilon ) \Big) \\2\pi \lambda_M r_m e^{-\pi\lambda_M r_m^2} dr_m\Big],
\end{multline}
where the expectation is with respect to the total interference from neighboring transmitters (i.e. MeNB, SeNB, or MC).

Using the Laplace transform for random variable i.e. $\mathcal{L}_X(s) = \mathbb{E}[e^{-sX}]$, and using the fact that $I_M$, $I_S$, and $I_{\tilde{a}}$ are independent random variables, $p_{BH}$ becomes:
\begin{equation}
\label{eq_shan}
\footnotesize = \int_{0}^\infty \mathcal{L}_{I_M}\Big(\frac{\theta r_m^{\alpha_i}}{P_M}\Big) \mathcal{L}_{I_S}\Big(\frac{\theta r_m^{\alpha_i}}{P_M}\Big)  \mathcal{L}_{I_{\tilde{a}}}\Big(\frac{\theta r_m^{\alpha_i}\gamma \epsilon}{P_M}\Big) 2\pi \lambda_M r_m e^{-\pi \lambda_M r_m^2}dr_m,
\end{equation}\normalsize

\subsection{Backhaul Success Probability : when $\kappa = 1$}
\label{app_BH_succ_prob}
Starting from Eq. \ref{eq_shan}, we find $\mathcal{L}_{I_M}(s)$ as \cite{andrews2011tractable}:
\begin{equation}
\label{eq_L_IC}
	\mathcal{L}_{I_M}(s) = \exp(-\pi r_m^2\lambda_M \rho(\theta, \alpha_i)),
\end{equation}
where $ \rho(\theta, \alpha_i) = \theta^{2/\alpha_i}\int\limits_{\theta^{-2/\alpha_i}}^{\infty}\frac{1}{1+\Lambda^{\alpha_{n}/2}} d\Lambda$.

Then, we find $\mathcal{L}_{I_S}(s)$ as:
\begin{equation}
	\mathcal{L}_{I_S}(s) = \exp\Big\{-\lambda_S \pi r_m^2 \beta(\alpha_i)\Big(\frac{P_S\theta}{P_M}\Big)^{2/\alpha_i}\Big\},
\end{equation}
where $\beta(\alpha) = \Gamma(\alpha/2)\Gamma(1-\alpha/2) = \frac{(2\pi/\alpha)}{\sin(2\pi/\alpha)}$ and $\Gamma(.)$ is the Gamma function.\\

Now considering $s = \gamma \epsilon \theta (r_m/r_{\tilde{a}m})^{\alpha_i} \frac{Po}{Pc}$, where $r_{\tilde{a}m}$ is the distance between the BH-antenna and AL-antenna for the MC and $h^\omega_{{\tilde{a}},v} \sim \exp(1)$, we can find $\mathcal{L}_{I_{\tilde{a}}}(s)$ for Eq. \ref{eq_shan} with the knowledge that $\mathcal{L}_h(s) = 1/(1+s)$ as:
\begin{equation}
\label{eq_L_Io}
\mathcal{L}_{I_{\tilde{a}}}(s) = \frac{1}{1 + \gamma \epsilon \theta (r_m/r_{\tilde{a}m})^{\alpha_i} P_{\tilde{a}}/Pc},
\end{equation}

Hence, $p_{BH}$ becomes:
\begin{multline}
\label{eq_shan2a}
p_{BH} = \int\limits_{r_m=0}^{\infty}\mathcal{L}_{I_{\tilde{a}}}(s) \mathcal{L}_{I_M}(s) \mathcal{L}_{I_S}(s) e^{-\lambda_M\pi r_m^2}2\lambda_M\pi r_m= \\\int\limits_{0}^{\infty} \exp\Bigg\{-\lambda_M\pi r_m^2\Big[\rho(\theta, \alpha_i) + \frac{\lambda_S}{\lambda_M}\frac{\pi}{2}\sqrt{\frac{P_S}{P_M}\theta} + 1\Big] \Big\} \\\times \frac{1}{1 + \frac{\theta \gamma \epsilon P_{\tilde{a}} r_m^{\alpha_i}}{r_{\tilde{a}m}^{\alpha_i}P_M\pi^2\lambda_M^2}} 2\pi\lambda_M r_mdr_m,
\end{multline}

Eq. \ref{eq_shan2a} does not have a closed-form expression. However, for $\alpha_i = 4$, a more tractable form can be obtained for numerical evaluation. Considering $\varpi = \frac{1}{1+ \lambda_M\pi r_m^2}$, Eq. \ref{eq_shan2a} can be transformed into Eq. \ref{eq_p1_0}. 

\subsection{Backhaul Success Probability : when $\kappa = 0$}
\label{app_success_BH_kappa_0}
The success probability $p_{BH}$ when $\kappa = 0$ is:
\begin{equation}
p_{BH} = 2\pi \lambda_M \int_{r_m>0} \mathcal{L}_{I_M}(s) \mathcal{L}_{I_{\tilde{a}}}(s) e^{-\pi \lambda_M r_m^2} r_m dr_m,
\end{equation}

Using $\mathcal{L}_{I_M}(s)$  in Eq. \ref{eq_L_IC} and $\mathcal{L}_{I_{\tilde{a}}}(s)$ in Eq. \ref{eq_L_Io},  $p_{BH}$ becomes:
\begin{equation}
p_{BH} = \exp(-\pi r_m^2\lambda_M \rho(\theta, \alpha)) \times \frac{2\pi\lambda_M \exp(-\pi \lambda_M r_m^2)}{1 + \gamma \epsilon \theta (r_m/r_{\tilde{a}m})^{\alpha_i} P_{\tilde{a}}/Pc} ,
\end{equation}
\begin{equation}
\label{eq_for_kappa0_appendix}
p_{BH}= 2\pi \lambda_M \int_{r_m>0}  \frac{e^{-\pi \lambda_M (1 + \rho(\theta, \alpha_i))r_m^2}}{1 + \gamma \epsilon \theta (r_m/r_{\tilde{a}m})^{\alpha_i} P_{\tilde{a}}/Pc} r_m dr_m,
\end{equation}

Eq. \ref{eq_for_kappa0_appendix} is not a closed-form expression. However, for $\alpha_i = 4$, numerically tractable form for Eq. \ref{eq_for_kappa0_appendix} can be evaluated by considering $z = \frac{1}{1+r_m}$ which is represented as Eq. \ref{eq_p1_1}.\\
This completes the proof \qed  

\section{Downlink Cellular Link Success Probability}
\label{app_success_DL_CL}
Following Eq. \ref{eq_cellular_user}, the success probability for DL cellular transmission ($p_{DL}$) can be given as:

\begin{equation}
p_{DL} = \mathbb{P}\Big[h_{M,u_{\bar{i}}}^\omega > \frac{\theta r_{u_{\bar{i}}}^{\alpha_i}}{P_M}(I_M + I_S + I'_{\tilde{a}})\Big],
\end{equation}

Using CCDF of $h_{M,u_{\bar{i}}}^\omega$, we get:
\begin{equation}
\label{eq_p2_temp}
p_{DL} = \mathbb{E}\Big[\exp\Big\{-\frac{\theta r_{u_{\bar{i}}}^{\alpha_i}}{P_M}(I_M + I_S + I'_{\tilde{a}})\Big\}\Big],
\end{equation}
where expectation is with respect to the random variables $I_M, I_S,$ and $I'_{\tilde{a}}$.
Following independence of $I_M, I_S,$ and $I'_{\tilde{a}}$ and the techniques used in Appendix \ref{app_success_DL_BH} and following $\alpha_i = 4$, we can obtain $p_{DL}$ by using following Laplace transforms with Eq. \ref{eq_p2_temp},

$$\mathcal{L}_{I'_M}(s) = \exp\Big\{\frac{-\pi r_{u_{\bar{i}}}^2 \lambda_M}{2}\sqrt{\frac{P_M\theta}{P_M}}\Big\},$$ $$\mathcal{L}_{I'_S}(s) = \exp\Big\{\frac{-\pi r_{u_{\bar{i}}}^2 \lambda_S \kappa}{2}\sqrt{\frac{P_S\theta}{P_M}}\Big\},$$ $$\mathcal{L}_{I'_{\tilde{a}}}(s) = \frac{1}{1 + \frac{\theta}{P_M}\Big(\frac{r_{u_{\bar{i}}}}{r_{m u_{\bar{i}}}}\Big)^2},$$

Hence, $p_{DL}$ is given as Eq. \ref{eq_p3_final}. 

This completes the proof \qed

\section{Downlink Access Link Success Probability}
\label{app_success_AL}
We start by stating that $p_{AL}$ is the probability of success for MC-to-MUE link which is given as 

\begin{equation}
\label{eq_ref_0}
p_{AL} = \mathbb{E}_{I'_{\tilde{a}}} \Bigg[ \mathbb{P}\Big[h^\omega_{{\tilde{a}},v} > \theta \frac{I_C r_{\tilde{a}m}^{\alpha_o} \gamma \epsilon}{P_{\tilde{a}}}\Big] \Bigg],
\end{equation}

Let \large $I'_{\tilde{a}} = \frac{I_C r_{\tilde{a}v}^{\alpha_o} \gamma \epsilon}{P_{\tilde{a}}}$, \normalsize then 

\begin{equation*}
p_{AL} = \mathbb{E}_{I'_{\tilde{a}}} \Bigg[\mathbb{P}\Big[h^\omega_{{\tilde{a}},v} > \theta I'_{\tilde{a}}\Big] \Bigg],
\end{equation*}

Then the Success probability can be found as:

\begin{equation}\label{eq_p2_intermediate}
p_{AL} = \mathbb{E}_{I'_{\tilde{a}}} \Big[1 - \mathbb{P}\Big[h^\omega_{{\tilde{a}},v} \leq \theta I'_{\tilde{a}}\Big] \Big],
\end{equation}

As mentioned in Section \ref{sec_sys_model}, the transmission between AL-antenna and MUE inside MC has strong LOS component due to presence of directional antennas. Hence, the link between transmitter $\tilde{a}$ and receiver $v$ follows Rician fading. Consequently, the channel $h^\omega_{{\tilde{a}},v}$ will follow non-central Chi-squared ($\chi^2$) distribution. The PDF for $f_{h^\omega_{{\tilde{a}},v}}(h_{\tilde{a}})$ can be given as $\big[$Ch:3, \cite{goldsmith2005wireless}$\big]$:
\begin{equation}
\label{eq_Rician}
f_{h^\omega_{{\tilde{a}},v}}(h_{\tilde{a}}) = \frac{K+1}{P_{avg}} e^{ \frac{-KP_{avg} - (K+1)h_{\tilde{a}}}{P_{avg}}}
I_0\Big(2\sqrt{\frac{K(K+1)h_{\tilde{a}}}{P_{avg}}}\Big),
\end{equation}
where $K$ is the ratio of power for dominant to the scattered component of access link and $I_0(.)$ is the modified Bessel function of the first kind of Zeroth order. 

The K-factor determines how strong is the impact of the LOS component of the signal. For example, $K = 0$ means the signal follows multipath fading with no dominant LOS component. On the other hand, $K = \infty$ means that a direct LOS component eliminating all scattering waves. The average received power by Rician fading is $P_{avg} = \int\limits_{0}^{\infty} h_{\tilde{a}} f_{h^\omega_{{\tilde{a}},v}}(h_{\tilde{a}})dh_{\tilde{a}} = 2\sigma^2(K+1)$ \cite{goldsmith2005wireless}. If the scattered component of the link is modeled as the Gaussian random variable with the variance $\sigma^2 = 1/2$, then $P_{avg} = K+1$ \cite{goldsmith2005wireless} . Hence, Eq. \ref{eq_Rician} will becomes:
\begin{equation}
\label{eq_pdf_g}
f_{h^\omega_{o_i,j}}(h_{\tilde{a}}) = \frac{I_0(2\sqrt{Kh_{\tilde{a}}})}{e^{Kh_{\tilde{a}}}} .
\end{equation}

Following Eq. \ref{eq_p2_intermediate}, $\mathbb{P}\Big[h^\omega_{{\tilde{a}},v} \leq \theta I'_{\tilde{a}}\Big]$ is the CDF for the random variable $h^\omega_{{\tilde{a}},v}$ given as $F_h^\omega$. But since it is difficult to find the CDF of $h_{\tilde{a}}$ due to the presence of Zeroth order Bessel function, we will do mathematical manipulations using the expansion series provided in 8.447.1 in \cite{gradshteyn2014table}. Then the PDF of $h^\omega_{{\tilde{a}},v}$ becomes \cite{peng2014device}:
\begin{equation}
\label{eq_PDF_Rician}
f_{h^\omega_{{\tilde{a}},v}}(h_{\tilde{a}}) = \sum\limits^{\infty}_{j=0} \frac{(Kh_{\tilde{a}})^j}{e^{(K+h_{\tilde{a}})}(j!)^2}
\end{equation}

The CDF for Eq. \ref{eq_PDF_Rician} is given as:

\begin{equation}
F_{h^\omega_{{\tilde{a}},v}}(h_{\tilde{a}}) = \int\limits_{0}^{x}\sum\limits^{\infty}_{j=0} \frac{(Kh_{\tilde{a}})^j}{e^{(K+h_{\tilde{a}})}(j!)^2} h_{\tilde{a}} dh_{\tilde{a}}
\end{equation}

\begin{equation}
F_{h^\omega_{{\tilde{a}},v}}(h_{\tilde{a}}) = \sum\limits^{\infty}_{j=0} \frac{(K)^je^{-K}}{(j!)^2} \int\limits_{0}^{x}h_{\tilde{a}}^je^{-h_{\tilde{a}}} dh_{\tilde{a}}
\end{equation}

So $p_{AL}$ will become:
\begin{equation}
p_{AL} =  \sum\limits_{j=0}^{\infty}\sum\limits_{m=0}^{j} \frac{e^{K}}{K^j j!(j-m)!}(\theta)^n \eth(\theta,n)
\end{equation}
where
\begin{equation}
\label{equal_1}
\eth(\theta,n) = \int\limits_{0}^{\infty}e^{-y} y^n  f_{I'_{\tilde{a}}}(y) dy = (-1)^n D^n \mathcal{L}_{I'_{\tilde{a}}}(\theta)
\end{equation}
where $D^n(.)$ is the $n_{th}$ derivate of the function. The combined Laplace transform of $I'_{\tilde{a}}$ can be written as \cite{peng2014device}:
\begin{multline}
\mathcal{L}_{I'_{\tilde{a}}} = \exp \Bigg\{-\pi (\theta \gamma \epsilon r_{\tilde{a}v}^{\alpha_o})^{2/\alpha_i} \\ .\Bigg(\lambda_M\Big(\frac{P_M}{P_{\tilde{a}}}\Big)^{2/\alpha_i} + \kappa\lambda_S\Big(\frac{P_S}{P_{\tilde{a}}}\Big)^{2/\alpha_i}\Bigg) \beta(\alpha_i) \Bigg\}
\end{multline}
where $\beta(\alpha)$ is defined in Appendix \ref{app_success_DL_BH}.

We find $\eth(\theta,n) = (-1)^n\frac{d^n \mathcal{L}_{I'_{\tilde{a}}}(\theta)}{d\theta^n}$ as 

\begin{equation}
\label{eq_eth}
\eth(\theta,n)  = (-1)^n\frac{d^n \exp(-\Omega_{\kappa} \theta^{2/\alpha_i})}{d\theta^n} 
\end{equation}
where $\Omega_{\kappa}$ is given in Theorem 3.

Let $f(\theta) = \exp(-\Omega_{\kappa} \theta^{2/\alpha_i})$. Using $e^{(.)} = \sum\limits_{q=0}^{\infty}\frac{(.)^q}{q!}$, we can solve Eq .\ref{eq_eth} as: 

\begin{equation*}
\eth(\theta,n)  = (-1)^n\sum_{q=0}^{\infty}\frac{(-1)^{q} \Omega_{\kappa}^q}{q!} \theta^{\frac{2q}{\alpha_i} - n}\frac{\Gamma(2q/\alpha_i + 1)}{\Gamma(2q/\alpha_i -n + 1)}. 
\end{equation*}

So, $p_{AL}$ will become: 
\begin{multline}
\label{eq_p2_final_app}
p_{AL} = \sum\limits_{j=0}^{\infty} \sum\limits_{m=0}^{j} \frac{K^j (-\theta)^{j-m}}{e^K j!(j-m)!} \sum\limits_{q=1}^{\infty} \frac{(-1)^q \Omega_\kappa^q}{q!}\frac{1}{P_{\tilde{a}}^q}\theta^{\frac{2q}{\alpha_i} - (j-m)}\\\frac{\Gamma(\frac{2q}{\alpha_i} + 1)}{\Gamma(\frac{2q}{\alpha_i} - (j-m) + 1)}. 
\end{multline}

Although Eq. \ref{eq_p2_final_app} is a closed form expression. However for large values of $j$ and $q$, $\frac{1}{j!}$ and $\frac{1}{q!}$ will reach zero. Hence we can determine the upper limit for index parameters $j$ and $q$ as $J$ and $Q$, respectively. Note that $J,Q$ must satisfy the condition that $\frac{1}{J!} \to 0$ and $\frac{1}{Q!}\to0$. Hence, we can rewrite the Eq. \ref{eq_p2_final_app} as Eq. \ref{eq_p2_final2}.

This completes the proof.   \qed

\section{Ergodic Rates for Shared Links}
\label{app_rate_DLBH}

\subsection{Ergodic Rate for Backhaul-link}

The Ergodic rate for BH-link of a typical MC $m$ is given as \cite{andrews2011tractable}:
\begin{equation}
T_{BH} = \int\limits_{0}^{\infty}\mathbb{E}[\ln(1 + \Upsilon_1(\omega,M\to m))] f(r_m) dr_m,
\end{equation}
where the expectations is with respect to the randomness of MC location around A-MeNB and the fading channel. For a positive random variable $\zeta$, $\mathbb{E}[\zeta] = \int\limits_{\tau>0} = \mathbb{P}(\zeta > \tau) d\tau$.

\begin{equation}
T_{BH} = \int\limits_{0}^{\infty}\int\limits_{0}^{\infty} \mathbb{E}\Big[\ln(1 + \frac{P_M r_m^{-\alpha_i}h^\omega_{M,m}}{I_M + I_S + I_{\tilde{a}} \gamma \epsilon})\Big] d\tau dr_m,
\end{equation}

For $\alpha_i = 4$, along with assumption that  $\bar{\iota} = \lambda_M \pi r_m^2$, we can write rate as:
\begin{multline}
\label{eq_Rates_2}
T_{BH} = \int\limits_{\bar{\iota}=0}^{\infty} \int\limits_{\tau=0}^{\infty} \exp \bigg\{ \bar{\iota}\bigg[1 + \sqrt{e^\tau-1} \Big(\frac{\pi}{2} - \tan^{-1}\frac{1}{\sqrt{e^\tau-1}} +\\ \frac{\kappa}{2}\frac{\lambda_S}{\lambda_M}\sqrt{\frac{P_S}{P_M}}\Big)\bigg]\bigg\} \times \frac{1}{1 + \frac{(e^\tau-1)\mathcal{F}\bar{\iota}^2}{\pi^2\lambda_M^2}} \ d\tau\ d\bar{\iota},
\end{multline}
where \large$\mathcal{F} = \frac{P_{\tilde{a}} \gamma \epsilon}{P_M r_{\tilde{a}m}^4}$\normalsize. Since Eq. \ref{eq_Rates_2} is still combination of two improper integral and remains intractable in this form, we can solve it using numerical integration method. We consider $g = \frac{1}{1 + \tau}$ and $\sigma = \frac{1}{1 + \bar{\iota}}$ to get the form as shown in Eq. \ref{eq_Rates_3}. 

\subsection{Ergodic Rate for Access-Link}

The ergodic rate for AL $T_{AL}$ can be found as:
\begin{equation}
T_{AL} = \mathbb{E}[\ln(1+\Upsilon_3) > \psi],
\end{equation}
\begin{equation}
T_{AL} = \int\limits_{\psi>0} \mathbb{E}_{I_M}\Big[P[h_{\tilde{a}} > (e^\psi-1)\frac{\epsilon I_M r_{\tilde{a}v}^{\alpha_o}}{P_{\tilde{a}}}] \Big] d\psi,
\end{equation}
\begin{equation}
T_{AL} = \int\limits_{\psi>0} \mathbb{E}[1 - F_{h_{\tilde{a}}}(X)] d\psi,
\end{equation}
\begin{multline}
T_{AL} = \int_{\psi>0} \sum_{j=0}^{\infty}\sum_{m=0}^{j}\frac{K^j(1-e^\psi)^{j-m}}{e^K.j!(j-m)!}\\\sum_{q=0}^{\infty}\frac{(-1)^q\Omega^q}{q!}(e^\psi-1)^{2q/\alpha_i - (j-m)} \Psi_{(2q/\alpha_i,j-m)} d\psi,
\end{multline}
where $\Psi_{(2q/\alpha_i,j-m)}$ is given in Theorem 3.

\begin{multline}
T_{AL} = \int\limits_{\psi>0} \sum_{j=0}^{\infty}\sum_{m=0}^{j}\sum_{q=1}^{\infty}\frac{K^j(-1)^{j-m+q}(e^\psi-1)^{2q/\alpha_i}}{e^K.j!(j-m)!}\\\frac{\Omega^q}{q!} \Psi_{(2q/\alpha_i,j-m)}\ d\psi,
\end{multline} 

Considering $g = \frac{1}{1 + \psi}$, the ergodic rate for AL is represented as Eq. \ref{eq_Ergodicrate_AL}.

\subsection{Ergodic Rate for Cellular Downlink}

The Ergodic rate for cellular downlink transmission $T_{DL}$ can be given as:
\begin{equation}
T_{DL} = \mathbb{E}[\ln(1 + \Upsilon_2(\omega,M\to m))],
\end{equation}
where the expectation here is with respect to the fading effect. Following Appendix \ref{app_rate_DLBH}, $T_{DL}$ becomes

\begin{equation}
T_{DL} = \int_{t>0} \mathbb{P}\Big[ \Upsilon_2(\omega,M\to m) > e^t-1\Big] dt,
\end{equation}

\begin{equation}
T_{DL} = \int_{t>0} \mathbb{P}\Big[ h_{M,u_{\bar{i}}}^{\omega} > (e^t-1) \frac{r_{u_{\bar{i}}}^{\alpha_i}}{P_M} ({I_M + I_S + I'_{\tilde{a}}\epsilon}) \Big] dt,
\end{equation}

\begin{equation}
\label{eq_temp_R_DL}
T_{DL} = \int_{t>0} \mathbb{E}\Big[ \exp\Big(-\frac{(e^t-1)r_{u_{\bar{i}}}^{\alpha_i}}{P_M} ({I_M + I_S + I'_{\tilde{a}}\epsilon})\Big) \Big] dt,
\end{equation}
where the expectation is with respect to the random variables denoting interference. As shown in Appendix \ref{app_success_DL_BH}, Eq. \ref{eq_temp_R_DL} can be solved by finding following Laplace transform. 

$$\mathcal{L}_{I_M}(s) = \exp\Big\{\frac{-\pi r_{u_{\bar{i}}}^2 \lambda_M}{2}\sqrt{\frac{P_M(e^t-1)}{P_M}}\Big\},$$ 

$$\mathcal{L}_{I_S}(s) = \exp\Big\{\frac{-\pi r_{u_{\bar{i}}}^2 \lambda_S \kappa}{2}\sqrt{\frac{P_S(e^t-1)}{P_M}}\Big\},$$ 

$$\mathcal{L}_{I_{\tilde{a}}}(s) = \frac{1}{1 + \frac{(e^t-1)}{P_M}\Big(\frac{r_{u_{\bar{i}}}}{r_o}\Big)^2},$$

Following technique used in Appendix \ref{app_success_DL_BH}, we get $T_{DL}$ as:
\begin{equation}
T_{DL} = \int_{0}^\infty \frac{\exp\Big(-\frac{\pi r_{u_{\bar{i}}}^2}{2}\sqrt{e^t-1}\Big(\lambda_M + \lambda_S\kappa\sqrt{\frac{P_S}{P_M}}\Big) \Big)}{1 + \frac{e^t-1}{P_M}\Big(\frac{r_{u_{\bar{i}}}}{r_o}\Big)^2} dt,
\end{equation} 

Now considering $\tilde{g} = \frac{1}{1 + t}$, we can get Eq. \ref{eq_erg_cellular_DL} for cellular downlink ergodic rate which is solved through numerical integration method.

This completes the proof \qed

\bibliographystyle{IEEEtran}
\bibliography{Globecom2018_references}

\begin{thebibliography}{10}
\providecommand{\url}[1]{#1}
\csname url@samestyle\endcsname
\providecommand{\newblock}{\relax}
\providecommand{\bibinfo}[2]{#2}
\providecommand{\BIBentrySTDinterwordspacing}{\spaceskip=0pt\relax}
\providecommand{\BIBentryALTinterwordstretchfactor}{4}
\providecommand{\BIBentryALTinterwordspacing}{\spaceskip=\fontdimen2\font plus
\BIBentryALTinterwordstretchfactor\fontdimen3\font minus
  \fontdimen4\font\relax}
\providecommand{\BIBforeignlanguage}[2]{{%
\expandafter\ifx\csname l@#1\endcsname\relax
\typeout{** WARNING: IEEEtran.bst: No hyphenation pattern has been}%
\typeout{** loaded for the language `#1'. Using the pattern for}%
\typeout{** the default language instead.}%
\else
\language=\csname l@#1\endcsname
\fi
#2}}
\providecommand{\BIBdecl}{\relax}
\BIBdecl

\bibitem{andrews2014will}
J.~G. Andrews, S.~Buzzi, W.~Choi, S.~V. Hanly, A.~Lozano, A.~C. Soong, and
  J.~C. Zhang, ``{What will 5G be?}'' \emph{{IEEE Journal on selected areas in
  communications}}, vol.~32, no.~6, pp. 1065--1082, 2014.

\bibitem{gupta2015survey}
A.~Gupta and R.~K. Jha, ``{A survey of 5G network: Architecture and emerging
  technologies},'' \emph{{IEEE Access}}, vol.~3, pp. 1206--1232, 2015.

\bibitem{cimmino2014role}
A.~Cimmino, T.~Pecorella, R.~Fantacci, F.~Granelli, T.~F. Rahman, C.~Sacchi,
  C.~Carlini, and P.~Harsh, ``The role of small cell technology in future smart
  city applications,'' \emph{Transactions on Emerging Telecommunications
  Technologies}, vol.~25, no.~1, pp. 11--20, 2014.

\bibitem{jaffry2016making}
S.~Jaffry, S.~F. Hasan, and X.~Gui, ``Making a case for the moving small
  cells,'' in \emph{Telecommunication Networks and Applications Conference
  (ITNAC), 2016 26th International}.\hskip 1em plus 0.5em minus 0.4em\relax
  IEEE, 2016, pp. 249--251.

\bibitem{3GPP_spec}
``Technical specification group radio access network; mobile relay for evolved
  universal terrestrial radio access ({E-UTRA}),'' Third Generation Partnetship
  Project ({3GPP}), techreport TR 36.836, May 2012.

\bibitem{sui2012performance}
Y.~Sui, A.~Papadogiannis, W.~Yang, and T.~Svensson, ``{Performance Comparison
  of Fixed and Moving Relays under Co-Channel Interference},'' in
  \emph{{Globecom Workshops, 2012 IEEE}}, pp. 574--579.

\bibitem{shanPotenails}
S.~Jaffry, S.~F. Hasan, and X.~Gui, ``{Mobile Cells Assisting Future Cellular
  Communication},'' \emph{{IEEE Potentials (Accepted)}}.

\bibitem{tanghe2008evaluation}
E.~Tanghe, W.~Joseph, L.~Verloock, and L.~Martens, ``Evaluation of vehicle
  penetration loss at wireless communication frequencies,'' \emph{{IEEE
  Transactions on Vehicular Technology}}, 2008.

\bibitem{jaziri2016offloading}
A.~Jaziri, R.~Nasri, and T.~Chahed, ``Offloading traffic hotspots using moving
  small cells,'' in \emph{Communications (ICC), 2016 IEEE International
  Conference on}.\hskip 1em plus 0.5em minus 0.4em\relax IEEE, 2016, pp. 1--6.

\bibitem{yasuda2015study}
H.~Yasuda, A.~Kishida, J.~Shen, Y.~Morihiro, Y.~Morioka, S.~Suyama, A.~Yamada,
  Y.~Okumura, and T.~Asai, ``{A Study on Moving Cell in 5G Cellular System},''
  in \emph{{82nd IEEE Vehicular Technology Conference (Fall), 2015}}, 2015, pp.
  1--5.

\bibitem{sui2013energy}
Y.~Sui, A.~Papadogiannis, W.~Yang, and T.~Svensson, ``{The Energy Efficiency
  Potential of Moving and Fixed Relays for Vehicular Users},'' in \emph{78th
  IEEE Vehicular Technology Conference (Fall), 2013}, pp. 1--7.

\bibitem{jiang2015spatio}
F.~Jiang, K.~Thilakarathna, M.~A. Kaafar, F.~Rosenbaum, and A.~Seneviratne,
  ``{A Spatio-Temporal Analysis of Mobile Internet Traffic in Public
  Transportation Systems: A View of Web Browsing from the Bus},'' in
  \emph{Proceedings of the 10th ACM MobiCom Workshop on Challenged
  Networks}.\hskip 1em plus 0.5em minus 0.4em\relax ACM, 2015, pp. 37--42.

\bibitem{chae2012dynamic}
S.~Chae, M.~Z. Chowdhury, T.~Nguyen, and Y.~M. Jang, ``{A Dynamic Frequency
  Allocation Scheme for Moving Small-Cell Networks},'' in \emph{2012 Intl'
  Conference on ICT Convergence (ICTC)}, pp. 125--128.

\bibitem{jangsher2013resource}
S.~Jangsher and V.~O. Li, ``Resource allocation in cellular networks employing
  mobile femtocells with deterministic mobility,'' in \emph{Wireless
  Communications and Networking Conference (WCNC), 2013 IEEE}.\hskip 1em plus
  0.5em minus 0.4em\relax IEEE, 2013, pp. 819--824.

\bibitem{jangsher2015resource}
S.~Jangsher and V.~O.~K. Li, ``{Resource Allocation in Moving Small Cell
  Network},'' \emph{IEEE Trans. on Wireless Communications}, 2016.

\bibitem{jangsher2017backhaul}
S.~Jangsher and V.~O. Li, ``Backhaul resource allocation for existing and newly
  arrived moving small cells,'' \emph{IEEE Transactions on Vehicular
  Technology}, vol.~66, no.~4, pp. 3211--3219, 2017.

\bibitem{khan2017outage}
A.~Khan and A.~Jamalipour, ``An outage performance analysis with moving relays
  on suburban trains for uplink,'' \emph{IEEE Transactions on Vehicular
  Technology}, vol.~66, no.~5, pp. 3966--3975, 2017.

\bibitem{jaffry2018shared}
S.~Jaffry, S.~F. Hasan, and X.~Gui, ``Shared spectrum for mobile-cells backhaul
  and access link,'' \emph{arXiv preprint arXiv:1807.11619}, 2018.

\bibitem{sen2010successive}
S.~Sen, N.~Santhapuri, R.~R. Choudhury, and S.~Nelakuditi, ``Successive
  interference cancellation: A back-of-the-envelope perspective,'' in
  \emph{Proceedings of the 9th ACM SIGCOMM Workshop on Hot Topics in
  Networks}.\hskip 1em plus 0.5em minus 0.4em\relax ACM, 2010, p.~17.

\bibitem{mahmood2016analysing}
N.~H. Mahmood, I.~S. Ansari, G.~Berardinelli, P.~Mogensen, and K.~A. Qaraqe,
  ``Analysing self interference cancellation in full duplex radios,'' in
  \emph{Wireless Communications and Networking Conference (WCNC), 2016
  IEEE}.\hskip 1em plus 0.5em minus 0.4em\relax IEEE, 2016, pp. 1--6.

\bibitem{duarte2012experiment}
M.~Duarte, C.~Dick, and A.~Sabharwal, ``Experiment-driven characterization of
  full-duplex wireless systems,'' \emph{IEEE Transactions on Wireless
  Communications}, vol.~11, no.~12, pp. 4296--4307, 2012.

\bibitem{andrews2011tractable}
J.~G. Andrews, F.~Baccelli, and R.~K. Ganti, ``A tractable approach to coverage
  and rate in cellular networks,'' \emph{IEEE Transactions on Communications},
  vol.~59, no.~11, pp. 3122--3134, 2011.

\bibitem{contains2013hetnets}
I.~Contains and A.~Hedlund, ``Hetnets: Opportunities and challenges,''
  \emph{White paper-An Ascom Network Testing}, pp. 1--22, 2013.

\bibitem{lin2013towards}
X.~Lin, R.~K. Ganti, P.~J. Fleming, and J.~G. Andrews, ``{Towards Understanding
  the Fundamentals of Mobility in Cellular Networks},'' \emph{IEEE Transactions
  on Wireless Communications}, vol.~12, no.~4, 2013.

\bibitem{merwaday2016handover}
A.~Merwaday and I.~G{\"u}ven{\c{c}}, ``Handover count based velocity estimation
  and mobility state detection in dense hetnets,'' \emph{IEEE Transactions on
  Wireless Communications}, vol.~15, no.~7, pp. 4673--4688, 2016.

\bibitem{rohani2017improving}
B.~Rohani, K.~Takahashi, H.~Arai, Y.~Kimura, and T.~Ihara, ``{Improving Channel
  Capacity in Indoor 4x4 {MIMO} Base Station Utilizing Small Bi-Directional
  Antenna},'' \emph{IEEE Trans. on Antennas \& Propagation}, 2017.

\bibitem{ni2017self}
W.~Ni, ``A self-learning based antenna system for indoor wireless network,''
  \emph{International Journal of Advanced Pervasive and Ubiquitous Computing
  (IJAPUC)}, vol.~9, no.~4, pp. 78--87, 2017.

\bibitem{gong2010mobility}
Z.~Gong and M.~Haenggi, ``Mobility and fading: Two sides of the same coin,'' in
  \emph{Global Telecommunications Conference (GLOBECOM 2010), 2010 IEEE}.\hskip
  1em plus 0.5em minus 0.4em\relax IEEE, 2010, pp. 1--5.

\bibitem{molina2017lte}
R.~Molina-Masegosa and J.~Gozalvez, ``Lte-v for sidelink 5g v2x vehicular
  communications: A new 5g technology for short-range vehicle-to-everything
  communications,'' \emph{IEEE Vehicular Technology Magazine}, vol.~12, no.~4,
  pp. 30--39, 2017.

\bibitem{jaffry2018neighbourhood}
S.~Jaffry, S.~F. Hasan, and X.~Gui, ``{Neighbourhood-aware out-of-network D2D
  discovery},'' \emph{Electronics Letters}, vol.~54, no.~8, pp. 507--509, 2018.

\bibitem{goldsmith2005wireless}
A.~Goldsmith, \emph{{Wireless Communications}}.\hskip 1em plus 0.5em minus
  0.4em\relax {Cambridge Univ. Press}, '05.

\bibitem{gradshteyn2014table}
I.~S. Gradshteyn and I.~M. Ryzhik, \emph{Table of integrals, series, and
  products}.\hskip 1em plus 0.5em minus 0.4em\relax Academic press, 2014.

\bibitem{peng2014device}
M.~Peng, Y.~Li, T.~Q. Quek, and C.~Wang, ``Device-to-device underlaid cellular
  networks under rician fading channels,'' \emph{IEEE Transactions on Wireless
  Communications}, vol.~13, no.~8, pp. 4247--4259, 2014.

\end{thebibliography}

\end{document}